\documentclass[twoside,english]{elsarticle}
\usepackage[T1]{fontenc}
\usepackage[utf8]{inputenc}
\usepackage{amsmath}
\usepackage{amssymb}
\usepackage{graphicx}

\makeatletter

\journal{Computer Physics Communications}
\usepackage{lineno}

\makeatother

\usepackage{babel}
\begin{document}

\begin{frontmatter}{}

\title{Inverse Primitive Path Analysis}

\author[cs]{Carsten Svaneborg}

\ead{zqex@sdu.dk}

\cortext[cs]{Corresponding author}

\address[cs]{University of Southern Denmark, Campusvej 55, 5230 Odense M, Denmark}
\begin{abstract}
The primitive-path analysis (PPA) {[}R. Everaers et al. Science 303,
823, (2004){]} is an algorithm that transforms a model polymer melt
into its topologically equivalent mesh by removing excess contour
length stored in thermal fluctuations. Here we present an inverse
PPA algorithm that gradually reintroduces contour length in a PPA
mesh to produce an topologically equilvalent polymer melt. This enables
the generation of model polymer materials with well controlled topology.
As an illustration, we generate knitted model polymer materials with
a 2D cubic lattice of entanglement points using a synthetic PPA mesh
as a starting point. We also show how to combine PPA and inverse PPA
to accelerate stress relaxation approximately by an order of magnitude
in simulation time. This reduces the computational cost of computational
studies of structure-property relations for polymer materials.
\end{abstract}

\end{frontmatter}{}

\section{\label{sec:level1}Introduction}

The complex interplay between molecular topology and emergent properties
of soft-matter is of great interest for physics, biology, and chemistry.
The relation between viscoelasticity and topological entanglements
in linear polymer melts is well understood\citep{DoiEdwards86}. Entanglements
gives rise to localization of thermal fluctuations to tube like structures
along chains.\citep{Edwards_procphyssoc_67} Melt dynamics is explained
by the reptation motion of chains inside tubes with slow tube renewal
occuring at the chain ends.\citep{degennes71} Significant efforts
have been invested in expanding our understanding to non-concatenated
ring polymer melts. Ring polymers crumple and adapt random tree-like
structures due to intermolecular entanglements in dense solutions\citep{muller1996topological,grosberg2014annealed,rosa2014ring}
which gives rise to unexpected elastic properties\citep{kapnistos2008unexpected,ge2016self,huang2019unexpected}.

Grossberg et al. proposed that biological function relies on DNA being
in the form of crumbled unknotted globules, since entanglements and
knots would impose strong kinetic barriers on biological processes.\citep{grosberg1993crumpled}
Chromosome territories seen during interphase was proposed by Rosa
and Everaers to be a non-equilibrium steady state caused by topological
constraints between unentangled ring-like structures.\citep{rosa2008structure,rosa2010looping}
Kinetoplast DNA of Trypanosomes and other protozoan parasites has
been found to be a network of thousands of topologically connected
DNA loops. \citep{chen1995topology,klotz2020equilibrium}

Metal coordination complexes have been used as templates to chemically
synthesize catenanes, which are topologically linked ring polymers\citep{dietrich1983nouvelle}
and offer a route to the synthesis of polymer materials with a controlled
molecular topology such as periodic arrays of entanglements\citep{thompson1964reactions,busch1992structural,hubin2000template}.
Chains of concatenated loops have also been synthesized \citep{wu2017poly,datta2020self}
as well as three-foil knots \citep{dietrich1989synthetic,segawa2019topological}.
Polymer networks formed by covalent cross-linking of linear chains
are permanent, however, dynamic network structures can e.g. be synthesized
with woven cross-links that can reversibly turn into entanglements.\citep{li2022robust}
For recent advances in these fields see e.g. Refs. \citep{orlandini2021topological,zhang2022molecular,ashbridge2022knotting}.

Many properties of polymer materials can be studied with computationally
efficient coarse-grained polymer models.\citep{binder1995monte,kremer2000computer,MultiscalePeterKremerFaradayDiss2010}
The most popular generic Molecular Dynamics model is the bead-spring
model of Kremer and Grest (KG) \citep{grest1986molecular,kremer90,grest90a}.
Several computational studies exists where the KG model has been used
to study e.g. ring polymers\citep{rosa2008structure,rosa2010looping,halverson2011molecular,rosa2014ring,schram2019local},
blends between linear and ring polymers\citep{halverson2012,michieletto2020dynamical},
polycatenanes\citep{tubiana2022circular}, interphase chromosomes\citep{rosa2008structure,rosa2010looping},
and force-extension curves of different knots\citep{caraglio2015stretching}.
Recently we used the KG polymer models to assess the contribution
of topological entanglements and cross-links to the elastic modulus
of model PDMS rubbers.\citep{Gula2020Entanglement} 

The topological state of a KG polymer system can easily be characterized
by primitive path analysis (PPA).\citep{PPA,sukumaran2005identifying}
During PPA excess length stored in thermal fluctuations is removed
as chains contract to their primitive paths. The contraction process
conserves topological constraints due to neighboring chains. The degree
of contour length contraction of the resulting PPA mesh can used to
estimate the plateau modulus.\citep{PPA,Svaneborg2020Characterization}
Alternative algorithms such as Z/Z1/Z1+ code \citep{kroger2005shortest,kroger2023z1plus},
CReTA\citep{tzoumanekas2006topological} and others\citep{shanbhag2005chain}
have been suggested. These variations differ from the original algorithm
by minimizing contour length of piecewise linear curves rather than
tension of the bead-spring chains. Typically the original PPA algorithm
provides an estimate of the degree of contour length contraction,
which can be used to estimate the entanglement modulus, while the
other algorithms provides an estimate for the contour length between
entanglement points along chains. While the meshes they produce look
qualitatively similar, they differ by about a factor of two in their
estimation of the plateau modulus.\citep{zhou2005primitive,everaers2012topological}

To systematically investigate the effect of topology on the properties
of a polymer material, it is highly desirable to be able to design
model materials with specific topological states. Current algorithms
for generation of model polymer melts\citep{zhang2014equilibration,moreira2015direct,SvaneborgEquilibration2016,SvaneborgEquilibration2022}
do not preserve topology. This is because they rely on a push-off
process to minimize bead overlap, which allows chains to slip through
each other. Here we present an inverse primitive path algorithm (iPPA)
which addresses this problem. iPPA gradually reintroduces excess length
in a mesh by a continuous transformation between the PPA and KG force
fields. The result of the iPPA is the conversion of a synthetic mesh
into a topologically equivalent KG model material. We illustrate this
process by creating periodic plain-knitted KG polymer melts, designed
in such a way that entanglements form a regular 2D cubic lattice.
We also illustrate how stress relaxation can be accelerated by applying
a deformation not to a KG melt, but to its mesh, and then after a
fast mesh relaxation (energy minimization), we use iPPA to convert
the mesh back to a KG melt state. The resulting melt conformation
is computationally much cheaper to brute force equilibrate compared
to brute force equilibration of a deformed KG melt. 

In Sect. \ref{subsec:rPPA}, we introduce the force field switching
process that connects the PPA and KG force fields. Our approach to
monitoring topology is introduced in Sect. \ref{subsec:Topology-check},
and in Sect. \ref{sec:Reversible-PPA-protocol} we introduce the protocol
we propose for switching between PPA and iPPA force fields. Validation
of topology preservation is presented in Sect. \ref{sec:Characterization}.
In Sect. \ref{subsec:Synthetic-PPA-meshes}, we show how to generate
a mesh with a controlled topology and perform iPPA to generate a KG
melt with the same topology. We show how iPPA can accelerate stress
relaxation in Sect. \ref{sec:Stress-relaxation}. We finish with a
conclusion in Sect. \ref{sec:Conclusions}. In an Appendix, we present
details of the KG model, and classical PPA force field. We also present
the details of how to use our iPPA force field and topology checking
code with the Large Atomic Molecular Massively Parallel Simulator
(LAMMPS)\citep{PlimptonLAMMPS,PlimptonLAMMPS2}.

\section{Methods\label{subsec:methods}}

\subsection{Switching PPA force field\label{subsec:rPPA}}

For the definition of the Kremer-Grest polymer force field, and the
standard PPA analysis method, we refer to the Appendix. In the original
PPA algorithm\citep{PPA} all intramolecular interactions are switched
off. This means that self-entanglements are lost. To preserve distant
self-entanglements along the chain, we have previously introduced
a PPA algorithm where intramolecular interactions are only switched
off within a specific window of chemical distances\citep{sukumaran2005identifying,Svaneborg2020Characterization}.
Here we gradually switch between the full KG force field and the windowed
PPA force field using the following potential for intra-molecular
pair interactions

\begin{equation}
U_{WCA}(i,j;\lambda)=\begin{cases}
0 & d(i,j)<w(\lambda)\\
U_{WCA,cap}(r,\delta(\lambda)r_{c}) & d(i,j)=w(\lambda)\\
U_{WCA}(r) & d(i,j)>w(\lambda)
\end{cases}.\label{eq:switchingU}
\end{equation}

Here $i,j$ denotes a pair of beads on the same chain. We assume beads
are numbered sequentially as $N_{min},\dots,N_{max}$ along this chain.
$d(i,j)$ denotes the chemical distance between the beads, that is
the number of bonds connecting the two beads. The chemical distance
is defined as $d(i,j)=|i-j|$ for linear chains, and $d(i,j)=\min\left(|i-j|,N_{max}-\max(i,j)-N_{min}+\min(i,j)+1\right)$
for cyclic chains. The size of the current window is denoted $w(\lambda)$,
and it depends on a control parameter $\lambda\in[0:1]$. The window
defines how interactions between bead pairs depend on the chemical
distance between them. Bead pairs within a chemical distance smaller
than the window do not interact. Bead pairs separated by a chemical
distance matching the current window interact via a force capped WCA
interaction. Finally, for beads further apart along the chain than
the current window we retain the full WCA interaction. This ensures
that distant self-entanglements are preserved.\citep{Svaneborg2020Characterization}

The force capped WCA potential defined by

\[
U_{WCA,cap}(r;r_{ci})=
\]

\[
\begin{cases}
\left|\frac{dU_{WCA}}{dr}\right|_{r=r_{ci}}(r-r_{ci})+U_{WCA}(r_{ci}) & \mbox{for \quad}r<r_{ci}\\
U_{WCA}(r) & \mbox{for \quad}r_{ci}\leq r
\end{cases},
\]
here $r_{ci}$ denotes an inner cut-off distance below which the WCA
potential is replaced by a linear extrapolation. The inner cut-off
can be related to the energy at overlap (which we refer to as the
force-cap) $U_{0}(\beta_{c})=U_{WCA,cap}(0,\beta_{c}r_{c})$ where
the reduced cutoff is $r_{ci}=\beta_{c}r_{c}$ with $\beta_{c}\in[0:1]$.
Then $U_{WCA,cap}(r;\beta_{c}r)\rightarrow0$ for $\text{\ensuremath{\beta_{c}}}\rightarrow1$
while $U_{WCA,cap}(r,\beta_{c}r_{c})\rightarrow U_{WCA}(r)$ for $\beta_{c}\rightarrow0$,
thus by gradually decreasing $\beta_{c}$, we increase the repulsion
between beads. The force cap is related to $\beta_{c}$ as

\begin{equation}
\frac{U_{0}(\beta_{c})}{\epsilon}=1+\frac{13}{\beta_{c}{}^{12}}-\frac{14}{\text{\ensuremath{\beta}}_{c}{}^{6}}\label{eq:u0}
\end{equation}
which can be inverted to provide $\beta_{c}$ and thus $r_{ci}$ as
function of the force cap as

\begin{equation}
\beta_{c}(U_{0}/\epsilon)=\frac{13^{1/6}}{(7+\sqrt{36+13U_{0}/\epsilon})^{1/6}}\label{eq:beta}
\end{equation}

\begin{figure}
\includegraphics[width=0.5\columnwidth]{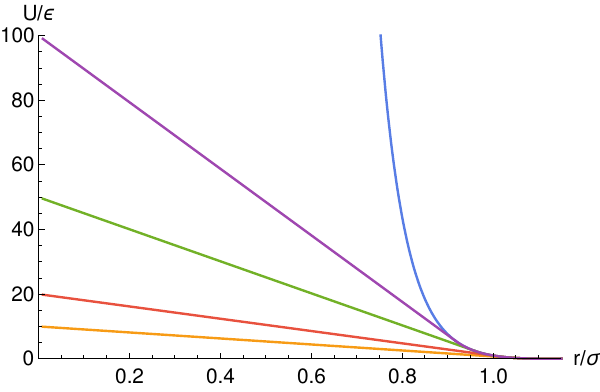}\caption{WCA potential (blue line) compared to force capped WCA potentials
for $U_{0}/\epsilon=10,20,50,100$ (yellow, red, green, respectively).\label{fig:WCA-potential}}
\end{figure}

Fig. \ref{fig:WCA-potential} shows the KG potential compared to force-capped
potentials with increasing overlap energies $U_{0}=10,20,50,100\epsilon$,
corresponding to decreasing cutoffs $\beta(10\epsilon^{-1})=0.9316$
down to $\beta(100\epsilon^{-1})=0.8175$.

To specify the switching protocol, we also need to define how the
window depends on $\lambda$. We use $w(\lambda)=\left\lfloor W(\lambda)\right\rfloor $
where $\left\lfloor x\right\rfloor $ denotes rounding down to nearest
integer (floor), and a continuous window $W(\lambda)$ function given
by

\[
W(\lambda)=\begin{cases}
W_{0} & \mbox{for \quad}\lambda=0\\
(1-\lambda)^{\alpha}W_{0}+1 & \mbox{for \quad}0<\lambda<1\\
0 & \mbox{for \quad}\lambda=1
\end{cases}.
\]

Where $W_{0}$ is the maximal size of the window, and $\alpha$ is
a parameter that controls the switching process. When $\lambda$ is
very small, $w(\lambda)=W_{0}$ and hence pair interactions are switched
off in the maximal window of chemical distances. As $\lambda$ increases
the window shrinks, when $\lambda$ is almost one, $w(\lambda)=1$
and the pair interaction between bonded beads is being switched. First
when $\lambda=1$ is the chemical window $w(1)=0$ in which case the
WCA pair interaction is applied between all beads. Within each integer
increment or decrement of the switching window, we also need to control
the force cap in Eq. (\ref{eq:switchingU}), this is done by choosing

\[
\beta(\lambda)=\beta_{c}+(1-\beta_{c})\left(W(\lambda)-w(\lambda)\right).
\]

The latter parenthesis is a sawtooth function. When it is zero, the
force cap within the current chemical window is at its maximal value
defined by $\beta(\lambda)=\beta_{c}$, while as the parenthesis approaches
unity, $\beta(\lambda)\rightarrow1$ and the potential is switched
off. Hence $\beta_{c}$ is a parameter that is defined by the maximal
force cap $U_{0}$ via eq. (\ref{eq:beta}). 

\begin{figure}
\includegraphics[width=0.5\columnwidth]{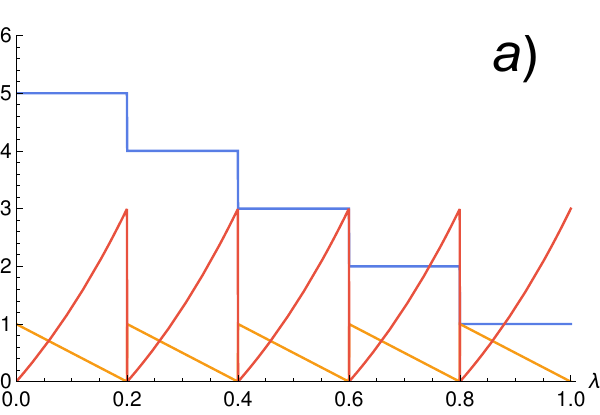}

\includegraphics[width=0.5\columnwidth]{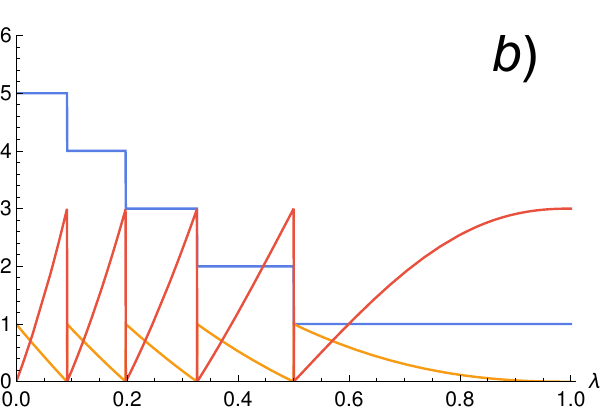}\caption{Switching functions $w(\lambda),$ $W(\lambda)-\left\lfloor W(\lambda)\right\rfloor ,$
and $U_{0}(\beta(\lambda))$ (blue, orange, red lines, respectively)
for $W_{0}=5,$$\beta_{c}$ corresponding to a force cap $U_{0}/\epsilon=3$
for $\alpha=1$(a) and $\alpha=\log W_{0}/\log2$ (b).\label{fig:Switching-functions-}}
\end{figure}

\begin{figure}
\includegraphics[width=0.25\columnwidth]{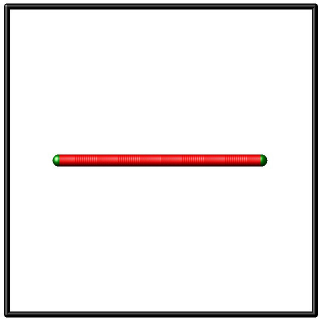}\includegraphics[width=0.25\columnwidth]{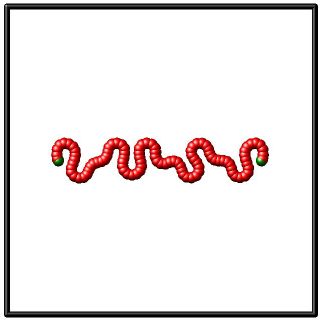}

\includegraphics[width=0.25\columnwidth]{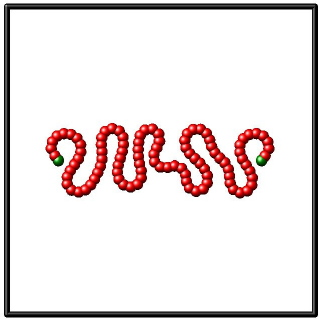}\includegraphics[width=0.25\columnwidth]{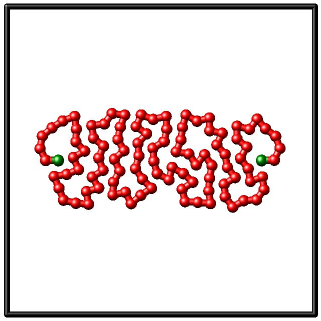}

\caption{Illustration of the push-off process for a 2D pinned mesh chain of
$100$ beads with a contour length contraction factor $\Lambda=0.15$
for $\lambda=0,0.25,0.5,1.0$ (top left to bottom right) for a protocol
with $W_{0}=5$, $\alpha=\log W_{0}/\log2,$and $U_{0}=200\epsilon$.\label{fig:Illustration-of-pushoff}}
\end{figure}

Fig. \ref{fig:Switching-functions-} shows two illustrative examples
of the force field switching protocol with $W_{0}=5$ and overlap
energy $U_{0}=3\epsilon$. The case $\alpha=1$ is shown in Fig. \ref{fig:Switching-functions-}a.
At $\lambda=0$, beads separated by $d<W_{0}=5$ bonds do not interact.
For beads separated by exactly $d=W_{0}$ bonds, $\delta(0)=1$ and
hence the pair interaction is also switched off. For beads separated
by $d>W_{0}$, full WCA interactions apply. This is consistent with
the PPA force field with a chemical window of $W_{0}$.

Increasing $\lambda$ in the interval $]0:0.2]$, $w(\lambda)=5$
and while the force cap grows from zero to the maximal value. This
has the effect of gradually introducing intramolecular excluded volume
interactions between all beads separated by exactly $5$ bonds. Thus
if any such bead pair is spatially overlapping, the force field transformation
will gently push them away from each other until they do not overlap.
At exactly $\lambda=0.2$ the force capped potential has reached its
maximal repulsion $U_{0}=3\epsilon$. When $\lambda$ is increased
above $0.2$, the force capped potential between beads 5 bonds apart
is switched from the maximal force capped potential to the full WCA
potential. Since all 5-distant beads have been pushed away from each
other, it is numerically safe to switch to the WCA potential. During
this process no changes were made to the intramolecular interactions
between beads separated by shorter or longer chemical distances. 

Increasing $\lambda$ in the interval $]0.2:0.4]$, repeats the process
but now with a chemical window $w(\lambda)=4$. Hence we progressively
introduce excluded volume interactions between bead pairs $4$ bonds
apart. Hence this process continues to introduce the full WCA interactions
between beads separated by fewer and fewer bonds by increasing the
relevant force capped potentials. When $\lambda$ is slightly above
$0.8$, we have $w(\lambda)=1$, and we start to introduce excluded
volume interactions between bonded beads, while the WCA interaction
is used for all beads separated by a larger chemical distance. When
$\lambda$ is just below unity, we have the maximal force capped repulsion
between bonded beads. When we reach $\lambda=1$, then $w(\lambda)=0$
and all intramolecular beads interact via the WCA potential. This
corresponds to the KG force field.

In the mesh, chains have rod-like conformations between entanglements
and due to contour length contraction many intramolecular beads will
overlap. During the iPPA push-off process, most of the contour length
is introduced when excluded volume is introduced between bonded beads.
Hence it makes sense to invest most of the computational effort during
the $w(\lambda)=1$ stage of the push-off process. Choosing $\alpha=\log W_{0}/\log2$
as in Fig. \ref{fig:Switching-functions-}b, we observe that the first
four switches occur in the interval $\lambda\in[0,0.5]$ while the
final switch for nearest neighbor beads occurs in the interval $\lambda\in]0.5:1]$.
Thus exactly half of the computer time is invested for the push-off
of bonded beads.

The process can also be run in reverse by reducing $\lambda$ from
one to zero. In this case, we start to remove excluded volume interactions
between nearest neighbors allowing them to overlap, and progressively
we remove interactions between beads up to the chosen maximal chemical
window $W_{0}$. As pair interactions are progressively switched off,
the bond interactions reels in the excess contour length, and pulls
the chain taut. This is a continuous variation of the standard PPA
contraction process where the force field is changed instantaneously.

Fig. \ref{fig:Illustration-of-pushoff} illustrates how a straight
PPA segment evolves during the inverse PPA using the protocol shown
in Fig. \ref{fig:Switching-functions-}b. Here the force cap was raised
to $U_{0}=200\epsilon$, which will be used in the simulations below.
The initial conformation corresponds to a single linear PPA chain
with the chains pinned in space. We observe that introducing excess
contour length creates wiggles along the initial chain-like conformation,
the wiggles grow in amplitude as the chemical window $w(\lambda)$
is reduced to one at $\lambda=0.5$. During the second half of the
simulation $\lambda\in[0.5,1]$ (bottom row) the target KG bond length
is established. Let the bond length of the PPA mesh be $l_{pp}=\Lambda l_{b}$,
where $l_{b}=0.965\sigma$ is the standard KG bond length, and $\Lambda$
denotes the PPA contour length contraction factor. For the standard
KG model $\Lambda(\kappa=0)=0.15$ (see the Appendix), in which case
the contour length grows by $\Lambda^{-1}=6.67$ during the iPPA push-off.

\subsection{Topology check\label{subsec:Topology-check}}

\begin{figure}
\includegraphics[width=0.5\columnwidth]{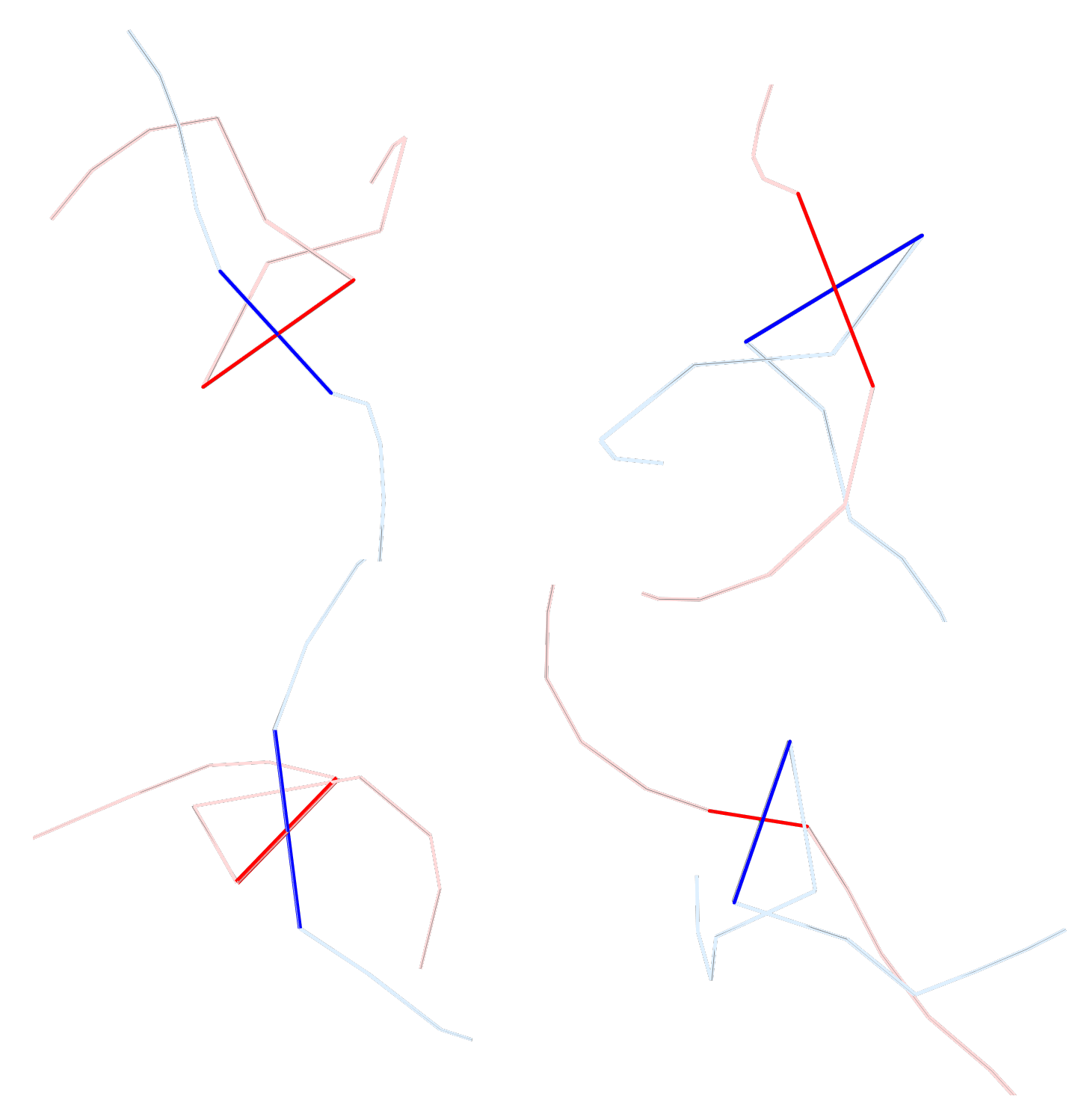}\caption{\label{fig:Visualizations-of-topology-violations}Visualizations of
the local chemical structure around a topology violation. Two segments
of the KG chains where only bonds are shown (pale red, pale blue).
The bond pair with the topology violation has been highlighted.}
\end{figure}

During the switching processes, we optionally can monitor for topology
violations. The classical approach is to monitor bond lengths and
stop the PPA algorithm if a bond exceeds a length of $1.2\sigma$.\citep{sukumaran2005identifying}
This approach works well for standard KG melts with stiffness $\kappa=0$,
but fails for stiffer melts where the contour length contracts much
less. Topology is not conserved, if a pair of bonds cross through
each other. Hence at each time step we check all bond pairs in close
spatial proximity, if they have crossed each other since the previous
time step.

The approach is based on the work of Sirk et al.\citep{sirk2012enhanced}
with some minor modifications. The algorithm works as follows: Each
bond defines a line segment, which can be extended into a line in
space. Based on the distance between the middle points of segment
pairs, we identify all pairs of line segments within a cutoff distance.
We choose cutoff $2\sigma$ since this is twice the bond length of
a KG model, and much larger than any PPA bond. Pairs of intra-molecular
bonds within a chemical distance less than the current switching window
are discarded, while all inter-molecular bond pairs and chemically
distant intra-molecular bond pairs are retained. To check a particular
pair of bonds, the code finds the two points defining the shortest
distance between the two lines. If one or both points are outside
the line segment, then the pair is discarded since the line segments
will be nearly parallel. Finally, the vector connecting the two shortest
distance points is calculated both for the current and for the previous
time step. If the dot product of these two vectors is negative, then
the shortest distance vector has flipped direction between the previous
time step to the current time step, which occurs when bonds cross.
This is counted as a topology violation.\citep{sirk2012enhanced}

Much of the code required for handling the spatial distribution of
line segments was already implemented in LAMMPS by Sirk et al. in
their work on segmental repulsive potentials (SRP) between bond pairs.\citep{sirk2012enhanced}
The code works by adding dummy particles at the middle of all bonds.
The dummy particles do not interact with other particles, but provide
a simple way of keeping track of spatial neighboring bonds using the
neighbor lists already implemented in LAMMPS. When LAMMPS evaluates
a SRP pair interaction between a pair of dummy particles, forces are
calculated based on the nearest distance between the two bonds corresponding
to the dummy particles, and the forces are then distributed between
the four beads defining the two bonds. Based on the SRP code, we developed
a LAMMPS pair style performing the topology check, and a LAMMPS fix
for storing necessary data. The two work in tandem, the pair style
compares all spatial neighboring bond pairs, the fix keeps a tally
of topology violations and also stores the bead coordinates of the
last time step.

We chose to implement the PPA force field and topology check as two
separate pair styles, since the latter is computationally costly,
and it might be useful in other circumstances. Having the ability
to mark all beads where topology violations occur, we can optionally
mark the beads involved in topology violations, and reset their positions
to their coordinates in the previous time step. This provides an approximate
way to correct for topology violations, however, we have not used
this option in the present paper. Instead we have implemented a polynomial
expansion of the FENE potential which allows topology violations to
be strongly suppressed. This is explained in more detail in the Appendix.
Optionally the topology checking code will save a file with the local
chemical environment around each topology violation it identifies.
Fig. \ref{fig:Visualizations-of-topology-violations} shows four such
topology violations. As discussed in Ref. \citep{sukumaran2005identifying},
the barrier transition state of the KG model is two perpendicular
bonds, where the topology violation occurs around the middle of each
bond. We observe that most topology violations occur where one chain
loop wraps around another chain that is under tension. The chain tension
causes a bond to extend to a bond length of approximately $1.2\sigma$,
while bonds in the loop are close to the equilibrium bond length.

\subsection{Protocol\label{sec:Reversible-PPA-protocol}}

\begin{table}
\begin{tabular}{|c|c|c|c|c|c|}
\hline 
Stage & Steps & $\Delta t$/$\tau$ & $\Gamma$/$m_{b}\tau^{-1}$ & $T$/$\epsilon$ & $k/\epsilon\sigma^{-2}$\tabularnewline
\hline 
\hline 
KG & - & $10^{-2}$ & $0.5$ & $1.0$ & $30$\tabularnewline
\hline 
Freezing & $10^{3}$ & $5\times10^{-3}$ & $50$ & $10^{-3}$ & $100$\tabularnewline
\hline 
PPA/iPPA & $10^{4}$ & $5\times10^{-3}$ & $50$ & $10^{-3}$ & $100$\tabularnewline
\hline 
Heating & $10^{3}$ & $10^{-4}$ & $200$ & $1.0$ & $30$\tabularnewline
\hline 
\end{tabular}

\caption{Overview of protocol\label{tab:Overview-of-protocol}}
\end{table}

The reversible PPA protocol depends on three parameters $W_{0},$$\alpha,$
and $U_{0}$. $W_{0}$ is the maximal size of the window of chemical
distances used when switching the force field. During primitive-path
analysis it should be chosen sufficiently large that the chains can
contract to their equilibrium PPA lengths. Here we use a constant
value of $W_{0}=10>\Lambda^{-1}(\kappa=0)=6.67$, since this is sufficient
for KG melts with $\kappa\geq0$. The maximal force capped potential
is defined by $U_{0}$. If the repulsion force is too small during
the push-off, then the simulation will crash when we switch to the
full WCA potential. We have observed that a typical value $U_{0}=200\epsilon$
is sufficient to ensure stability of simulations.

The $\alpha$ parameter allows us to control how many MD iterations
to spend within each switching window. $\alpha=1$ corresponds to
spending an equal number of integration steps within each chemical
distance step. However, most of the heat is generated during the push-off
of the nearest neighbors which occurs as $\lambda\rightarrow1$, hence
values $\alpha>1$ allows us to invest more integration steps during
this part of the push-off. For instance, if we aim to spend half of
the MD steps on the removing/introducing interactions within bonded
neighbors, then $w(\lambda)=1$ should occur at $W(\lambda=0.5)=2$.
The solution is to choose $\alpha=\log(W_{0})/\log(2)$. Fig. \ref{fig:Switching-functions-}b
shows the same force field switching, but with this choice of $\alpha$
parameter.

Before converting a melt state to a mesh with the force field transformation,
we freeze it to $T=0.001\epsilon$ using a short cooling stage with
a high friction. The force field transformation can be done within
$10^{4}$ steps, however, the resulting mesh state is not necessarily
converged. If necessary, we apply gradient descent minimization or
dampened Langevin dynamics with the $\lambda=0$ force field to converge
the mesh. To reverse the process, we perform an iPPA push-off again
using just $10^{4}$ steps. The entire force field switch occurs at
essentially zero temperature, and we use a short heat-up stage with
$T=1\epsilon$ and a high friction to thermalize the system before
running a KG simulation. The parameters are summarized in Tab. \ref{tab:Overview-of-protocol}.
As in the original PPA algorithm, we use a FENE potential with a spring
constant of $k=100\epsilon\sigma^{-2}$ during the force field switch.
Below we have also used a polynomial expansion of the FENE potential
to limit topology violations as described below. As a thermostat,
we choose Langevin dynamics.

\section{Results and Discussion}

\subsection{Characterization\label{sec:Characterization}}

\begin{figure}
\includegraphics[width=0.5\columnwidth]{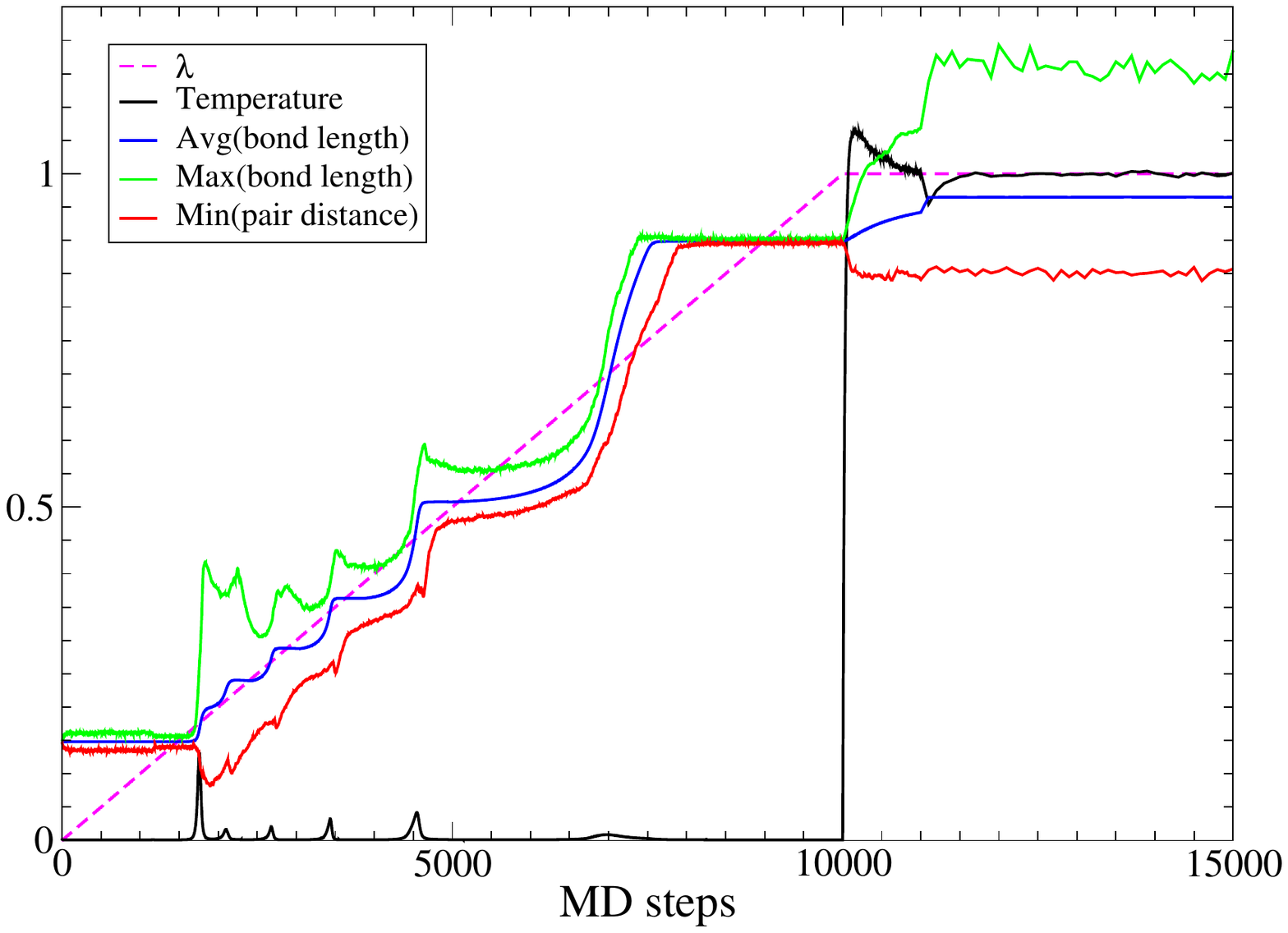}

\includegraphics[width=0.5\columnwidth]{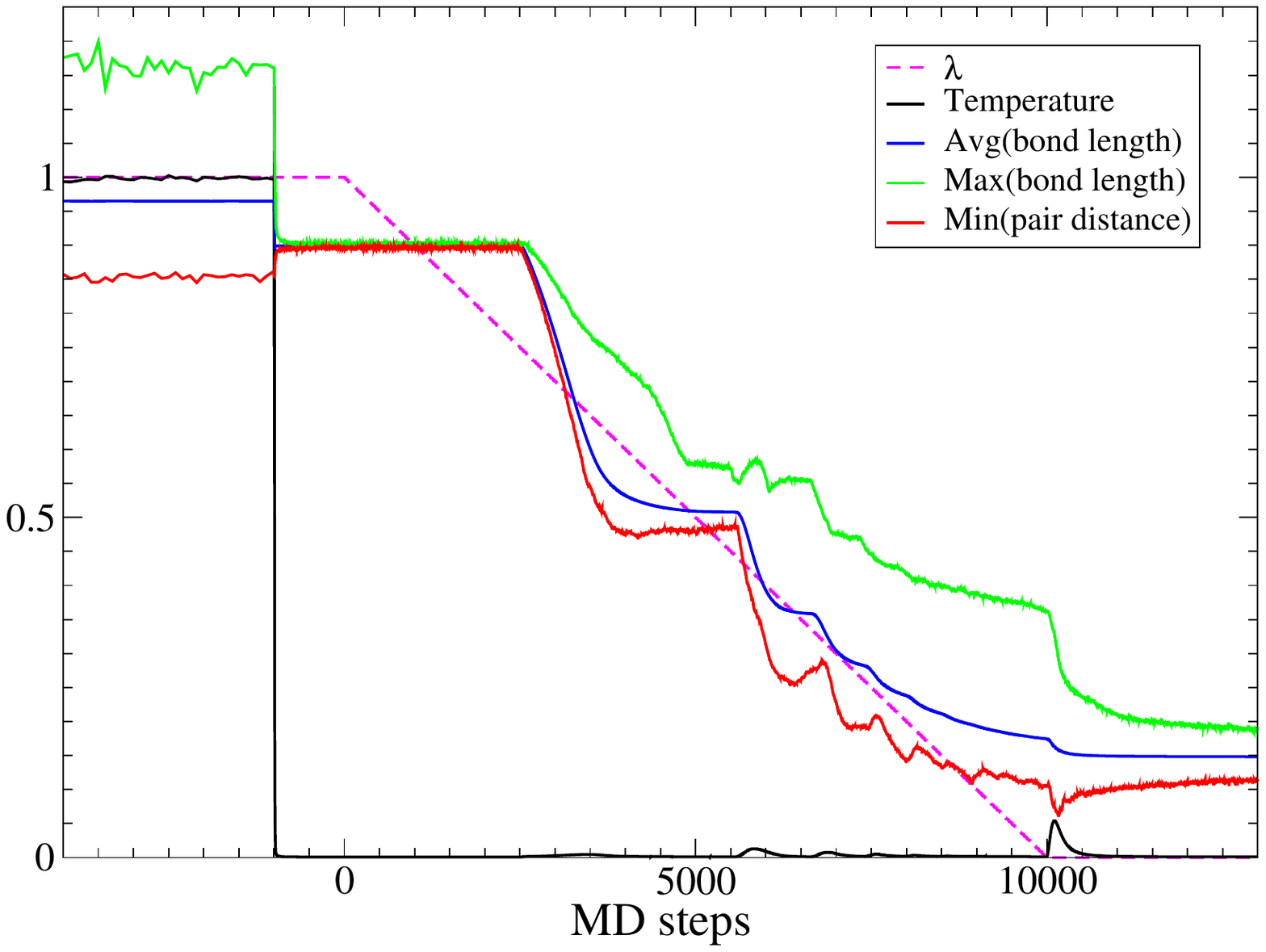}

\includegraphics[width=0.5\columnwidth]{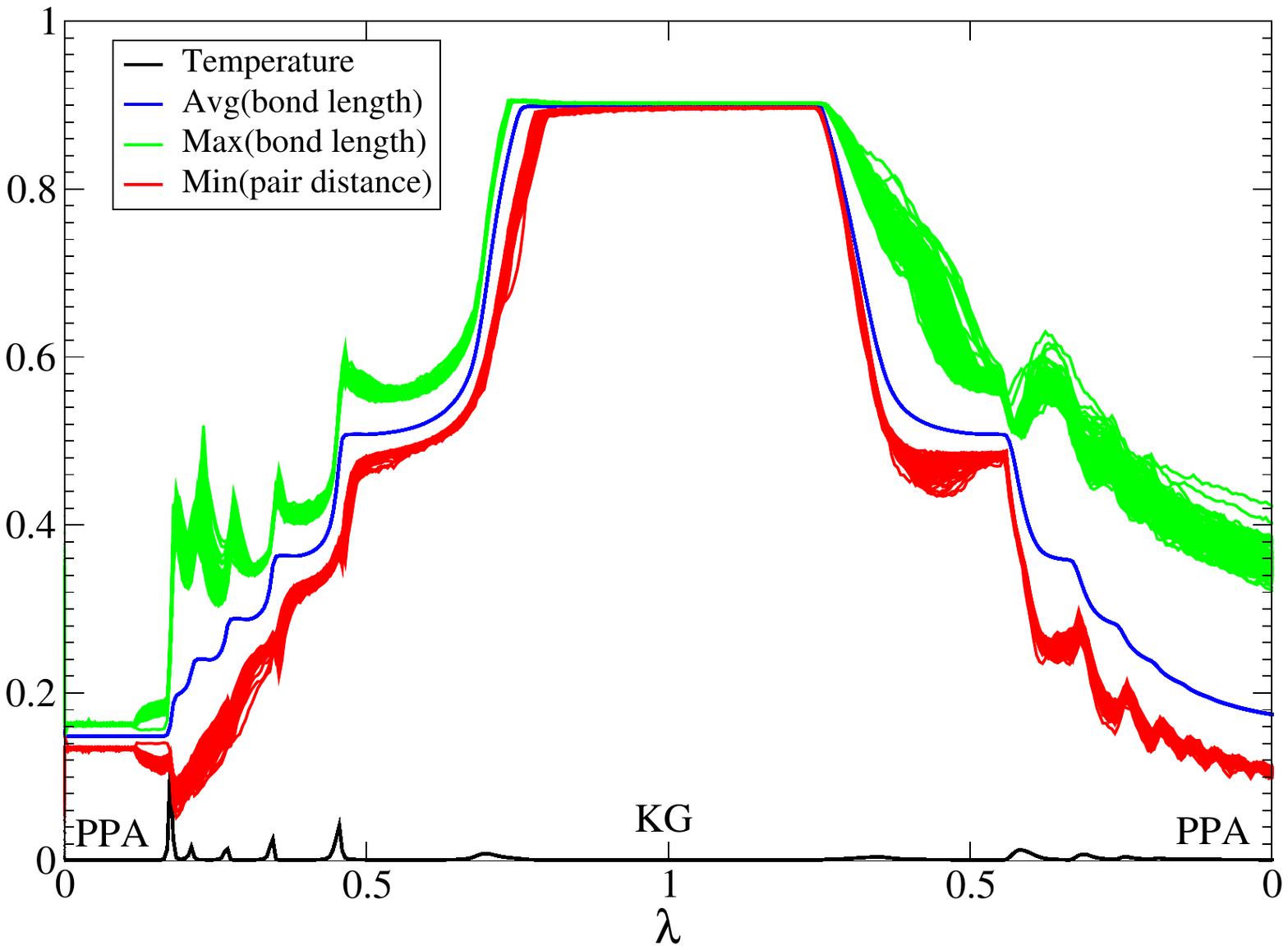}

\caption{Microscopic melt structure for the $\kappa=0$ system around the PPA
push-off (a) and contraction (b) during the first cycle. Also shown
is the results for all cycles superposed on each other (c).\label{fig:cycle}}
\end{figure}

\begin{figure*}
\includegraphics[width=0.25\textwidth]{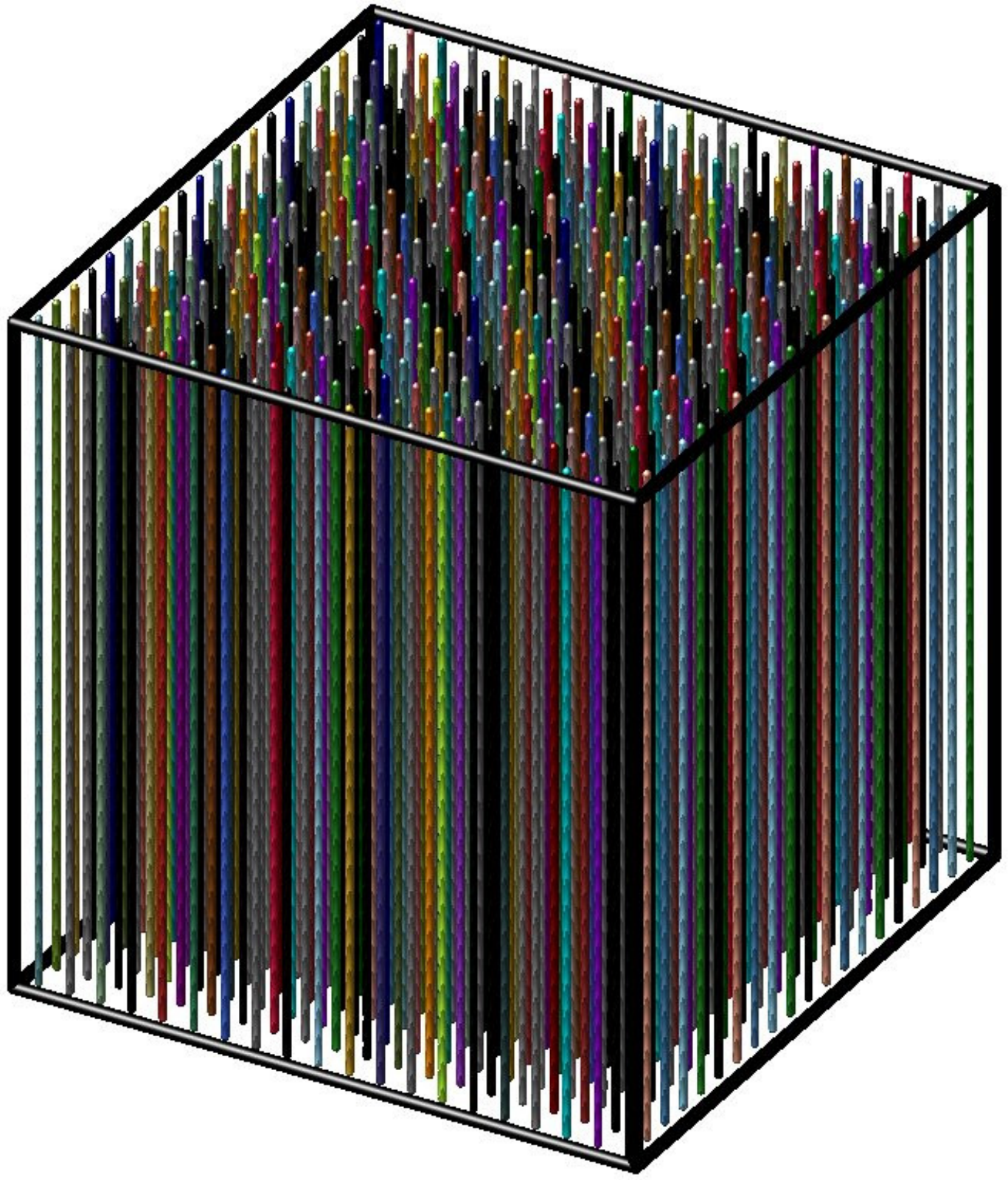}\includegraphics[width=0.25\textwidth]{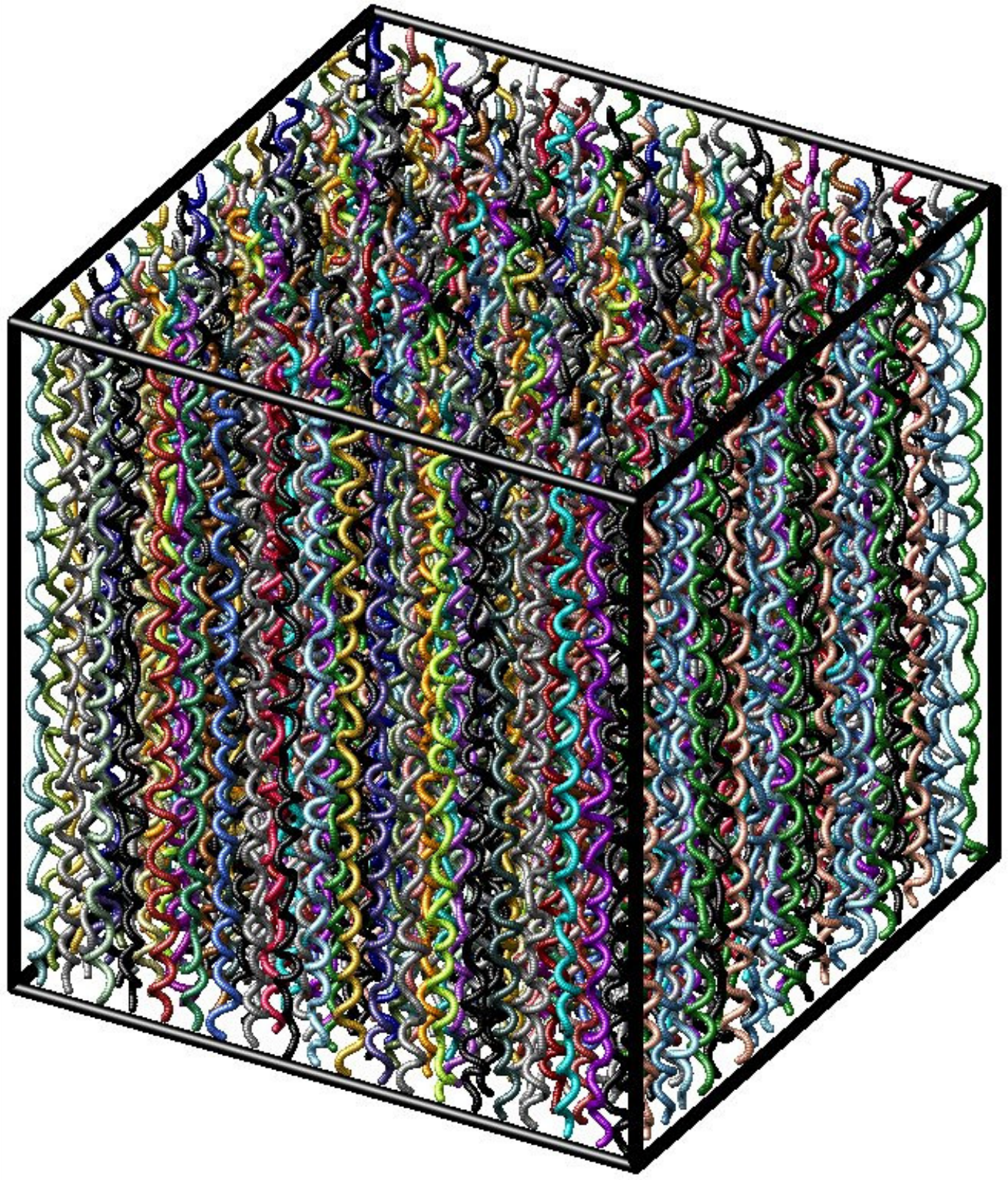}\includegraphics[width=0.25\textwidth]{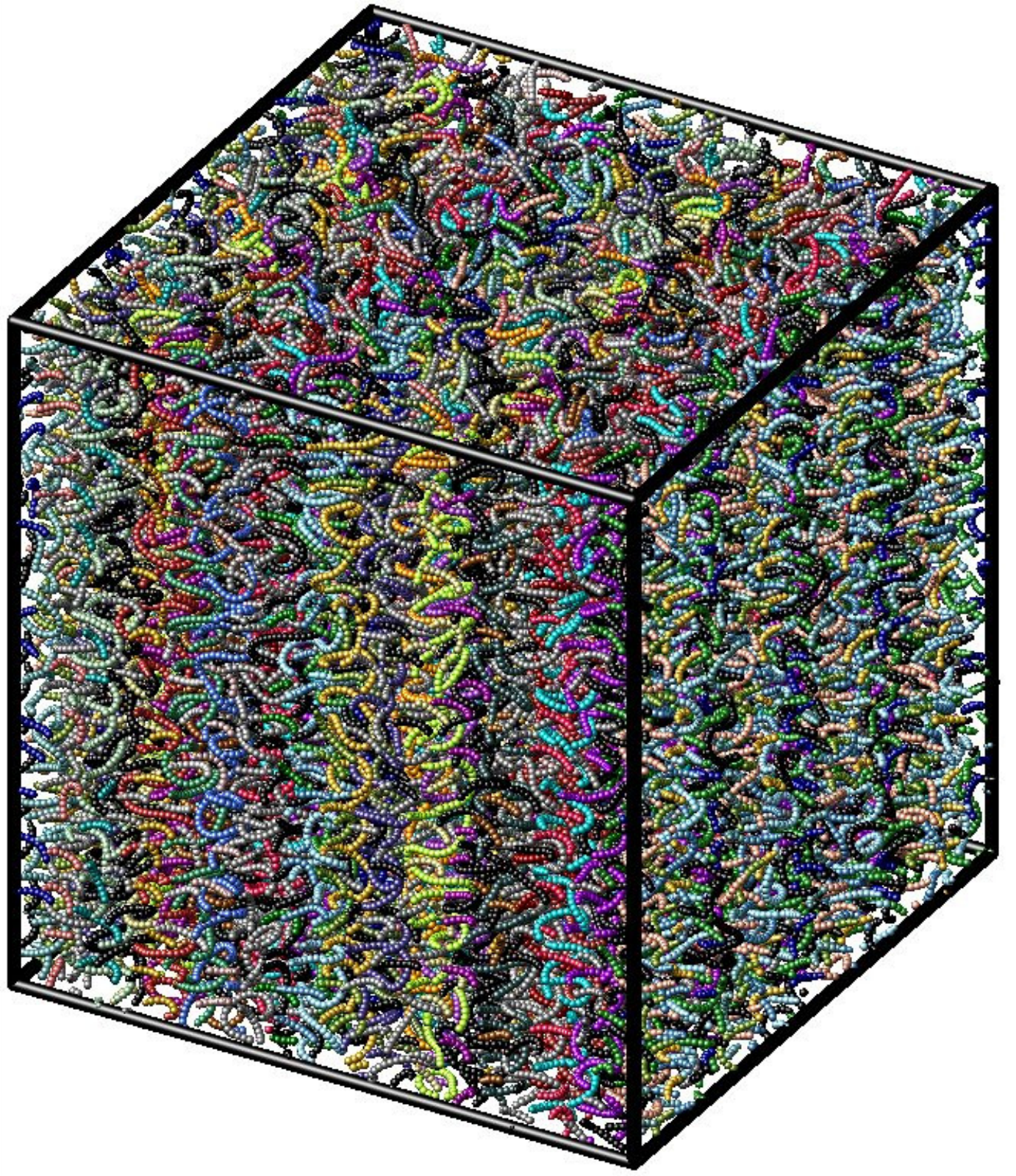}\includegraphics[width=0.25\textwidth]{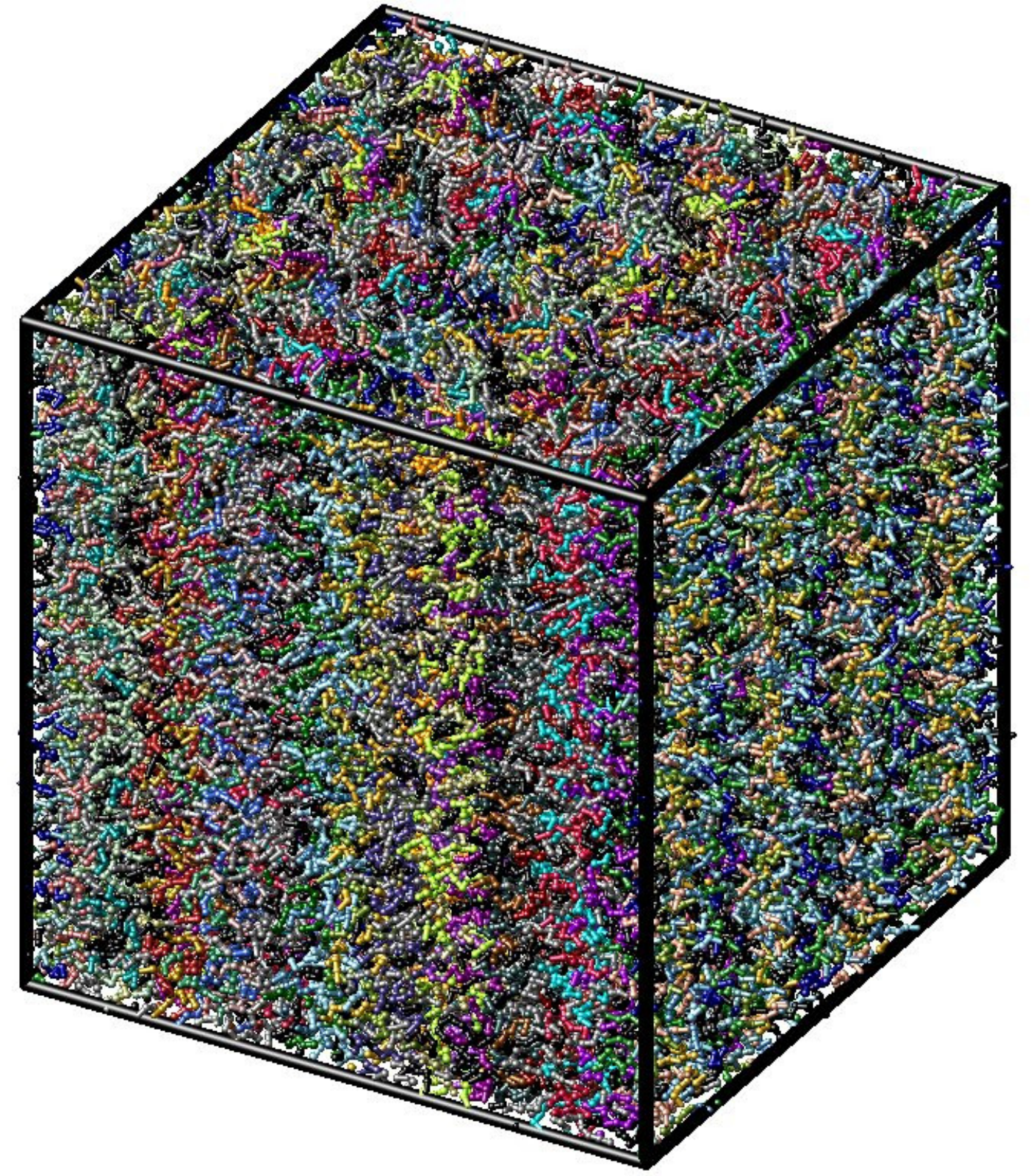}

\includegraphics[width=0.25\textwidth]{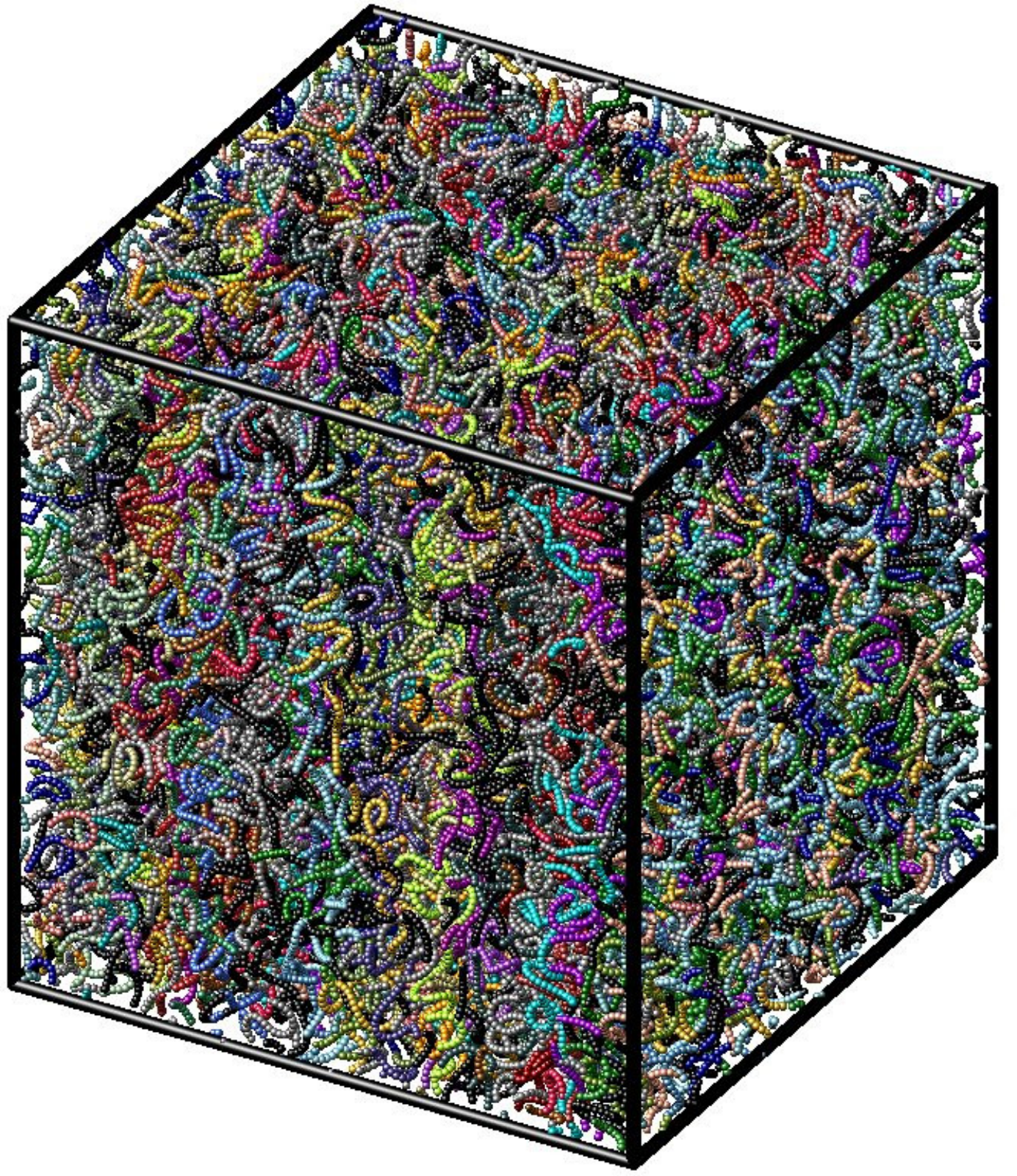}\includegraphics[width=0.25\textwidth]{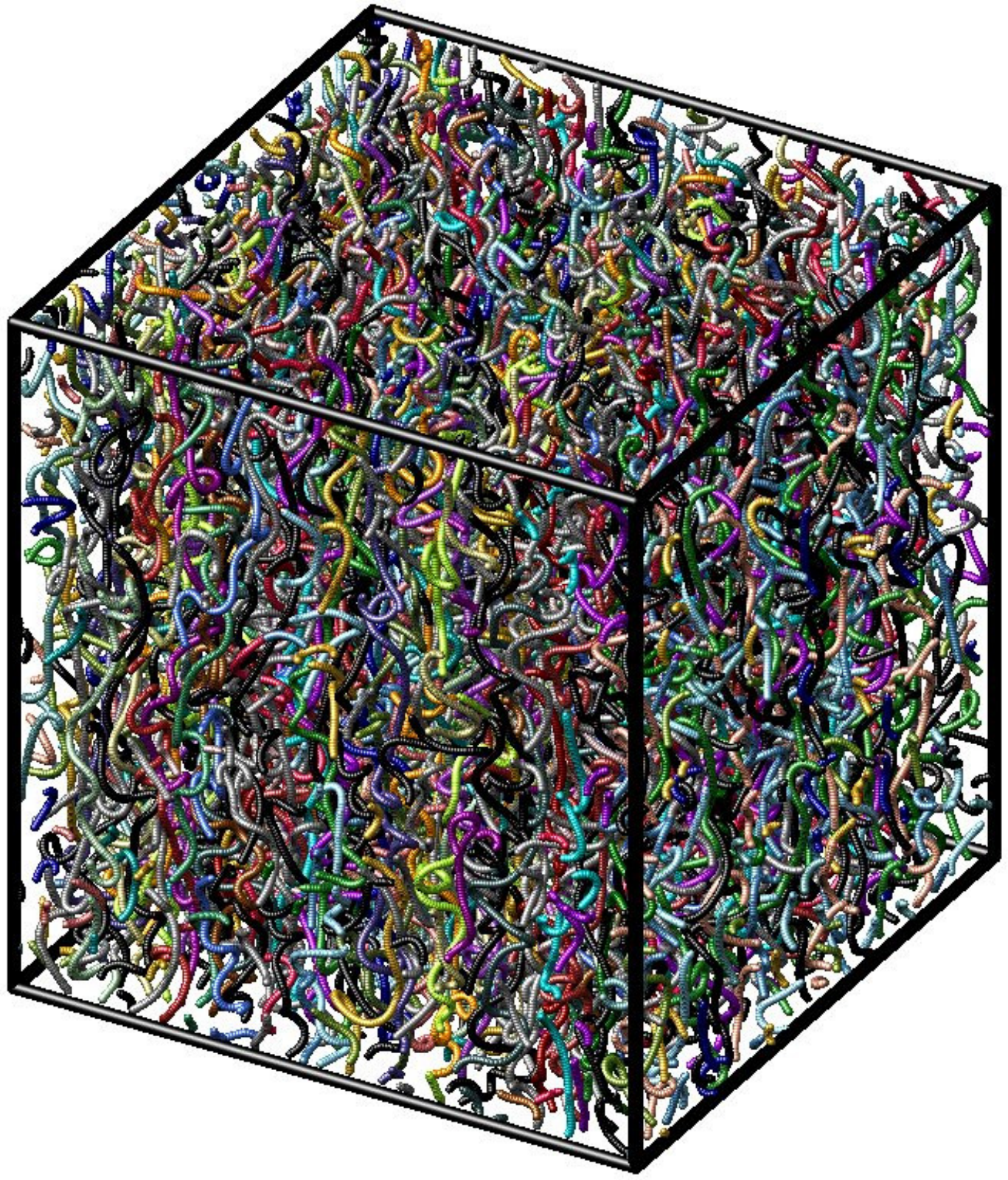}\includegraphics[width=0.25\textwidth]{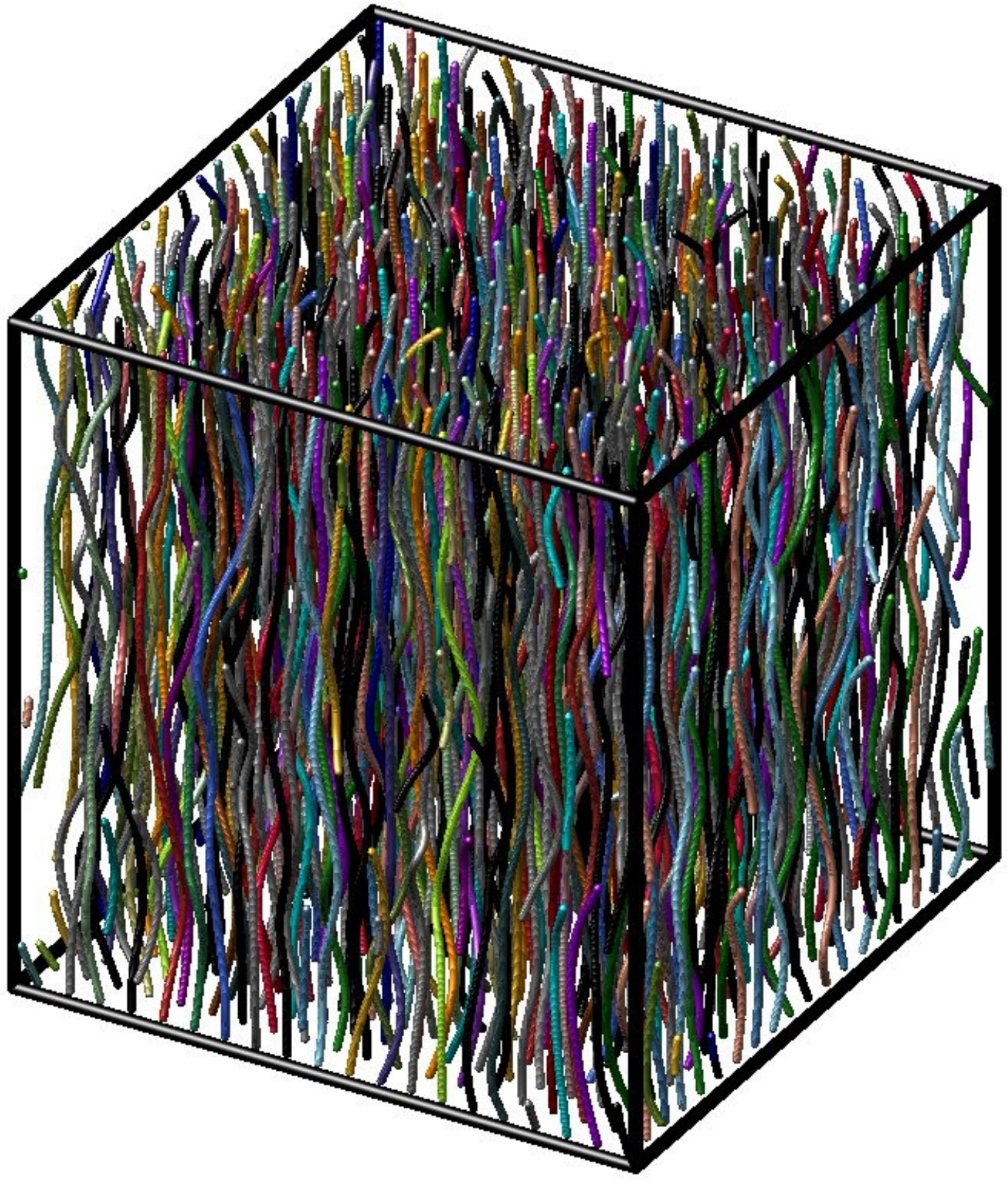}\includegraphics[width=0.25\textwidth]{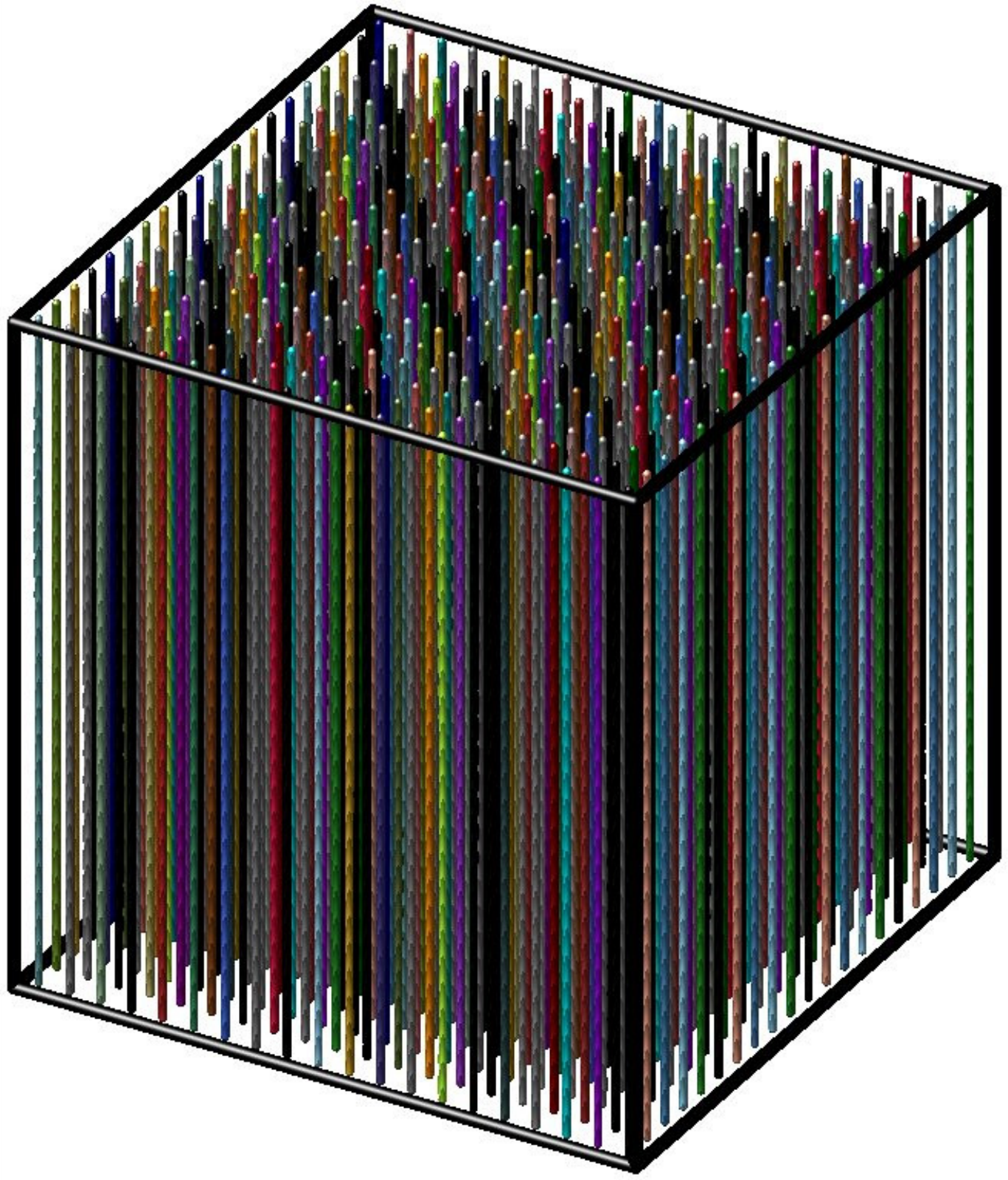}

\caption{Visualizations of a single iPPA-PPA cycle for a mesh corresponding
to a KG stiffness $\kappa=0$. The top row shows (left to right) conformations
during the iPPA push-off with $\lambda=0$, $0.25$, $0.5$, $1.0$.
The bottom row shows conformations during the PPA contraction for
$\lambda=0.5$, $0.25$, $0$, as well as the final conformation after
energy minimization. \label{fig:Visualizations-of-melts}}
\end{figure*}

To characterize the iPPA force field, we generate synthetic meshes
comprising $400$ linear chains of length $5a_{pp}(\kappa)$ for chain
stiffnesses $\kappa=-2,0,2,4$. Here $a_{pp}(\kappa)$ denotes the
Kuhn length of the primitive path mesh of a melt of linear chain,
for the corresponding numerical expressions, we refer to Refs. \citep{Svaneborg2020Characterization,SvaneborgEquilibration2022}
We choose the length of the simulation box to match the length of
the chains, while the lateral dimensions are chosen to reproduce the
KG bead density. Due to periodic boundary conditions, the two ends
of a chain are spatially adjacent, and we add a bond between the chain
ends turning the linear chains into straight loops. Finally, we pin
one of the chain ends in space. We run the KG simulation for $10^{5}$
MD steps after each iPPA stage before applying the next PPA stage.
After each cycle, we save the PPA mesh which directly enables us to
observe any topology violations. For each of the four systems, we
perform $100$ cycles of iPPA-KG-PPA transformations. Since we have
pinned a bead on each chain, this ensures that each iPPA-KG-PPA cycle
will result in the same mesh which makes it easy to visually detect
topology violations.

Fig. \ref{fig:cycle}a shows how the microscopic melt structure evolves
during the iPPA push off using the protocol defined above. The push-off
is performed between $0$ to $10000$ MD steps. We observe a gradual
monotonous increase of the minimum pair distance, while the maximal
bond distance also increases in a series of jumps followed by plateaus.
During the push-off the pair potential is gradually increased thus
increasing the energy of the system, the progressively larger repulsive
interactions cause beads to collide and produce heat which is removed
by the thermostat. We indirectly observe this as the increase of the
maximal bond distance which occurs concomitantly with spikes in the
temperature. The average bond length is bounded by the minimal pair
distance and maximal bond length, and we observe that the bond length
grows slowly during the push-off. Hence the net result of the iPPA
push-off is to introduce excess contour length in the chains. From
$10000$ to $11000$ MD steps we perform the short high-friction temperature
of the full KG force field, the system temperature rapidly grows to
the target temperature with a small overshoot. The high friction was
chosen to minimize this overshoot. Beyond $10100$ MD steps, we simulate
the full KG model with the standard friction.

Fig. \ref{fig:cycle}b shows how the microscopic melt structure evolves
during the gradual PPA contraction process. The contraction process
starts by freezing the melt using the KG force field with high friction
and zero temperature. The PPA contraction then proceeds by increasing
the spring constant, and reducing the force cap of pair interactions
between bonded beads. When the force cap gets sufficiently low bonds
start to contract, but reach a plateau due to the pair interaction
between beads two bonds apart. Eventually $\lambda$ is reduced enough
for this potential to drop, and the contraction process can continue.
At $\lambda=0$ we reach the end of the PPA contraction process. Finally,
we minimize the energy and see that both the average and maximal bond
length decreases to their equilibrium values. During the contraction
process we observe a single large temperature spike when the contraction
process begins, but otherwise no temperature spikes are observed. 

Fig. \ref{fig:cycle}c shows how all iPPA push-offs (left half of
the graph) and gradual PPA contractions (right half of the graph)
processes during the $100$ cycles superimposed on each other. We
observe a significant scatter of the maximal bond lengths, but the
dynamics is well controlled and the maximal bond value always stays
bounded. Similarly the minimum pair distance shows some scatter, but
we consistently reach large minimal pair distances ensuring stable
KG simulations. As expected, the average bond distance shows much
less scatter, and evolves in the same way during all $100$ cycles.
We observe similar behavior for the systems with other stiffnesses
(not shown). For all systems the maximal bond length is below $0.99\sigma$
during all $100$ PPA analysis cycles. We observe that the maximal
temperature of $1.38\epsilon$ during the heating stage and $0.20\epsilon$
during PPA analysis. During the simulations we did not observe a single
topology breakage event. 

Fig. \ref{fig:Visualizations-of-melts} shows visualizations of selected
conformations during the first cycle for KG models with stiffness
$\kappa=0$. The initial mesh is constructed as parallel straight
lines with a contour length contraction $\Lambda=0.15$. During the
PPA push-off contour length grows by a factor $\Lambda(\kappa=0)^{-1}=6.67$
which is introduced along the whole chain contour. Since the chain
is completely flexible, this causes wiggles along the contour of the
chain that progressively grow analogously to Fig. \ref{fig:Illustration-of-pushoff}.
In the final step of the PPA push-off the beads are densely packed
in space. After the long KG simulation, we apply a PPA contraction
that progressively removes excess contour length. In the final step
of the PPA contraction, we observe undulating roughly parallel chains.
After a final energy minimization we obtain a mesh of parallel straight
chain conformations identical to the starting mesh. To check for topology
violations, we visualized the meshes looking along the direction of
the chains, where any entanglements would be easy to observe. No topology
violations were observed.

Assuming the wiggles form a tube along the primitive path chain, then
we can estimate the tube radius as follows. Assuming a straight PPA
segment with $N$ beads, then it will have a length of $L_{pp}=\Lambda l_{b}N$.
Since the density of beads in the cylinder is $\rho_{b}=N/(\pi r^{2}L_{pp})$
we obtain the radius $r=(\pi\rho_{b}l_{b}\Lambda)^{-1/2}=(\pi\Lambda)^{-1/2}\xi=1.61\sigma$,
where $\xi=(\rho_{b}l_{b})^{-1/2}=1.10\sigma$ is the mesh size of
the KG model. The size of the wiggles is independent of the number
of beads in a segment and grows with an increased contour length contraction
factor as expected. In a mesh, the tubes containing wiggles will be
space filling, and we can estimate the volume of one wiggle as a cylinder
of length and radius r: $V_{w}=\pi r^{3}=13\sigma^{3}$ which corresponds
to $N_{w}=11$ beads per wiggle. Comparing that to the number of beads
between entanglements $N_{eb}(\kappa=0)=84$,\citep{Svaneborg2020Characterization}
we expect 7-$8$ wiggles to be created per entanglement segment for
standard KG systems with $\kappa=0$. For stiffer systems, the contour
length length will not expand as much as estimated above, hence the
wiggles will be smaller. For example for stiffer melts with $\kappa=2$,
the contour length contraction is just $\Lambda=0.41$ and thus the
wiggle radius $r=0.97\sigma$. A wiggle will contain about $N_{w}=2.5$
beads. Comparing that to the much smaller entanglement length of $N_{eb}(\kappa=2)=18$
beads, we still obtain $7-8$ wiggles per entanglement segment. Just
after the iPPA push-off, a melt will have the structure of local wiggles
along the primitive paths of the mesh. The large scale chain statistics
is the same as that of the PPA, which is the same as the original
precursor melt. The wiggles cause local chain perturbations in the
chain statistics. These perturbations are much smaller than the entanglement
length, and thus are expected to require a reequilibration simulation
that are significantly shorter than the entanglement time.

\subsection{Synthetic PPA meshes\label{subsec:Synthetic-PPA-meshes}}

\begin{figure}
\includegraphics[width=0.45\columnwidth]{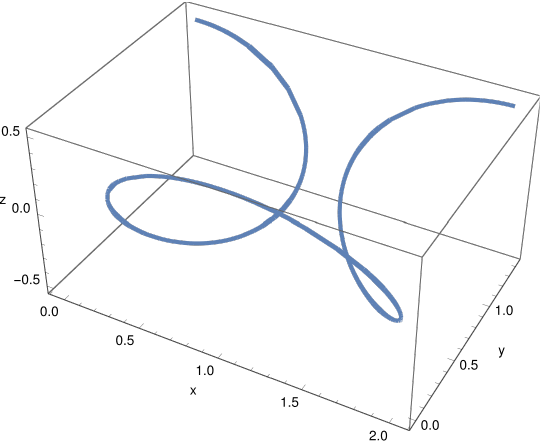}\caption{A view of the parametric curve $\gamma(t)$ for $t\in[0,1]$ with
$a=0.6$, $h=0$, $o=0$, and $s=1$.\label{fig:ParametricCurve}}
\end{figure}

\begin{figure}
\includegraphics[width=0.45\columnwidth]{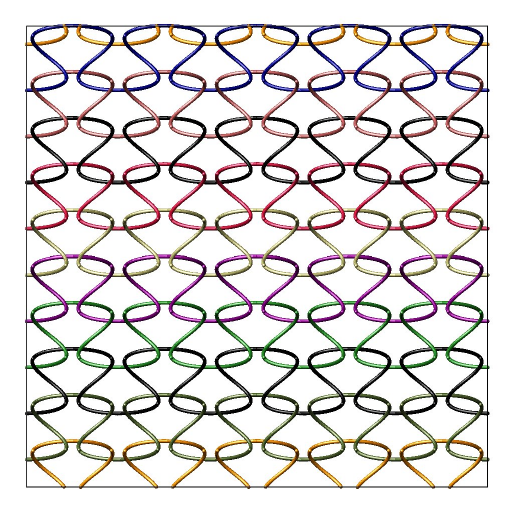}

\includegraphics[width=0.45\columnwidth]{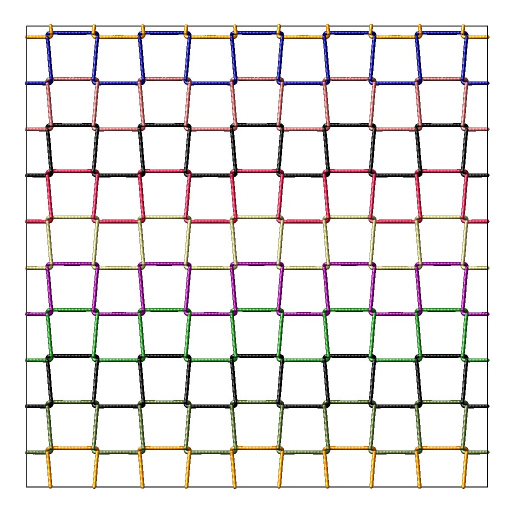}

\caption{Replicated parametric curves (left) and the corresponding primitive-path
mesh (right) for $n=5$. Parameters are $a=0.6$, $h=0.2$.\label{fig:RepeatedParametric}}
\end{figure}

\begin{figure*}
\includegraphics[width=0.33\textwidth]{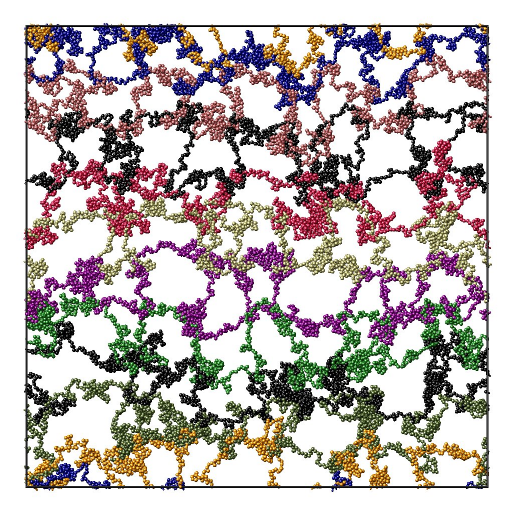}\includegraphics[width=0.33\textwidth]{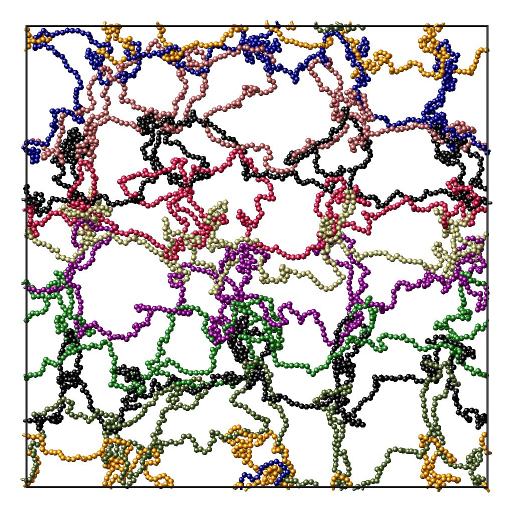}\includegraphics[width=0.33\textwidth]{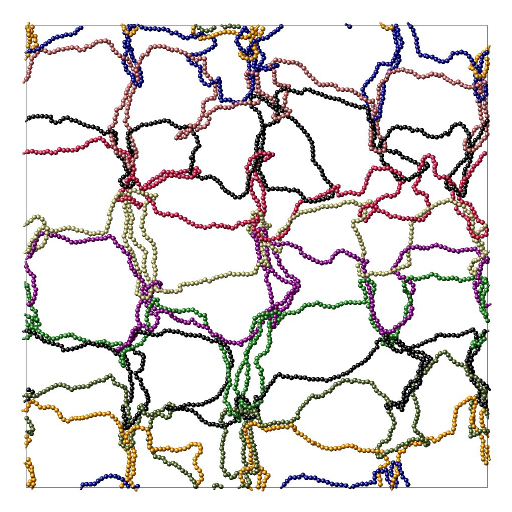}
\caption{Visualization of equilibrium KG melt conformations with 2D knitted
topology for stiffness values $\kappa=0,2,4$. Only chains in the
first layer are shown corresponding to Fig. \ref{fig:RepeatedParametric}
.\label{fig:yearnmelt}}
\end{figure*}
To illustrate how iPPA can generate KG melts with a controlled topology,
we choose to generate melts with a 2D knitted topology. Instead of
attempting to generate the primitive path of such a structure directly,
we start with the parametric curve recently proposed by K. Crane\citep{kcrane},
that produces the desired topology. Here we define the curve as

\begin{equation}
{\bf \gamma}(t)=s\left(\begin{array}{c}
2t+a\sin\left(4\pi t\right)\\
\left(0.5+h\right)\left[1-\cos\left(2\pi t\right)\right]+o\\
0.5\left[1-\cos\left(4\pi t\right)\right]
\end{array}\right)\label{eq:yarn}
\end{equation}

Fig. \ref{fig:ParametricCurve} shows the parametric curve eq. \ref{eq:yarn})
for $t\in[0,1]$. The curve defines a complicated 3D loop starting
at $(0,0,0)$ and ending at $(2,0,0)$. The curve does oscillations
along the three axes with the $y$ oscillation being half the frequency
of the $xz$ oscillations. The $a$ parameter defines the ``overshoot''
of the oscillation in comparison to the term describing linear motion.
We keep this parameter fixed at $a=0.6$. The $s$ parameter plays
the role of a global scale factor. For $o=h=0$, a single loop is
bounded by the volume $[0,2s]\times[0,s]\times[0,s]$.

The utility of this curve becomes apparent, when we choose $t\in[0,n]$
for some integer $n$, and replicate the curve $2n$ times with offsets
$o=0,1,2,\dots,2n-1$ along the $y$ axis. Fig. \ref{fig:RepeatedParametric}
(top) shows the structure generated for $n=5$ with periodic boundary
conditions applied. It has $5$ loops repeated along the horizontal
axis, and 10 repeated curves along the vertical axis. The $h$ parameter
defines the excess amplitude of the oscillation, we have chosen $h=0.2$
to ensure that each loop thread the loops of the neighboring curves.
We apply periodic boundary conditions, and note the system has the
volume $[0,2ns]\times[0,2ns]\times[0,s]$. This ensures that the top
and bottom curves thread each other via periodic boundary conditions.
Performing primitive-path analysis (\ref{fig:RepeatedParametric}
(bottom), we get a mesh that is a regular 2D cubic lattice. The mesh
has $2n$ entanglements in the horizontal and vertical directions,
thus in total it has $n{{}^2}$ regular spaced pair-wise entanglements,
corresponding to $2n$ entanglement points along each curve. Each
entanglement strand (straight segment between entanglements) has the
same length, and it carries the same tension, hence the stress carried
by the mesh in the horizontal and vertical directions is identical.
If this was not the case, then the mesh would be build with an anisotropic
tension. 

To generate KG model melts, we convert the parametric curves to a
bead-spring representation. Noting that the length of an entanglement
strand (a straight segment in the mesh) is exactly $s$, the most
natural choice is to identify this scale with the Kuhn length of the
primitive path $a_{pp}(\kappa)$, since this corresponds to the length
of a tube segment. The Kuhn length depends on the stiffness of the
target melt.\citep{Svaneborg2020Characterization} Here, we have chosen
to generate melts for $\kappa=0,2,4$, which corresponds to $a_{pp}(0)=12.3\sigma$,
$a_{pp}(2)=8.19\sigma$, and $a_{pp}(4)=9.65\sigma$, respectively.\citep{Svaneborg2020Characterization} 

To decorate the parametric curves with beads, we make use of the fact
the bond length of the PPA mesh is given by $\Lambda(\kappa)l_{b}$
(the contraction ratio $\Lambda(\kappa)$ is shown in Fig. \ref{fig:Contour-contraction}
for an equilibrium KG melt of linear chains). The final ingredient
required is to notice that a loop of the parametric curve has more
contour length ($7.7s$) than the PPA mesh ($4.4s)$. We would expect
$4s$, but the additional length is due to the finite size of the
beads. Hence to generate a KG melt, we decorate the parametric curve
with beads every contour length of $l_{pc}=7.4\text{\ensuremath{\Lambda}(\ensuremath{\kappa)}\ensuremath{l_{b}}/4.4}$
in which case the resulting PPA mesh will reproduce the desired mesh
bond length $l_{pp}(\kappa)$ of a linear melt with the same stiffness
within a $2\%$ error. The curve eq. (\ref{eq:yarn}) is not parametrized
by contour length, hence to find the next bead position ${\bf \gamma}(t_{next})$,
we use a Newton Raphson algorithm to find $t_{next}$ such that $|{\bf \gamma}(t_{next})-{\bf \gamma}(t)|=l_{pc}$.
For the present systems, each loop requires a decoration of $N_{b}(0)=705$,
$N_{b}(2)=185,$ and $N_{b}(4)=130$ beads, respectively. 

The bead density of the mesh should be the same as for a KG melts
$\rho_{b}=0.85\text{\ensuremath{\sigma^{-3}}}$. The procedure above
produces parametric curves with a density that is an order of magnitude
too low. Since each layer is pseudo-2D, the density can be fixed by
compressing the mesh along the $z$ direction to reproduce the target
density. For a single layer, this procedure will fail, since the resulting
system has a $z$ dimension that is less than the size of a bead.
This leads to spurious effects. However, this can be resolved by replicating
the $xy$ layer a number of times along the $z$ direction to ensure
the $z$ dimension of the system is much larger than the bead size.
We also decrease the oscillation in the $z$ direction by $10\%$
and apply a random shift in the $xy$ plane to each layer. To obtain
the final mesh, we apply a primitive-path analysis to an uncompressed
set of layers, followed by a compression of the resulting meshes along
the $z$ direction to reproduce the target density. Having an initial
mesh designed with the desired melt topology and bead density, we
proceed to use the iPPA protocol described in Sect. \ref{tab:Overview-of-protocol}
to introduce excess contour length. We then perform a simulation of
$5\times10^{6}$steps to thermalize and equilibrate the resulting
KG melts. This is long compared to the entanglement times of the systems.\citep{Svaneborg2020Characterization}

Fig. \ref{fig:yearnmelt} shows the visualizations of the resulting
melt conformations. The topology remains clearly visible. We observe
that thermal fluctuations give rise to random walk-like conformations
for the flexible melt ($\kappa=0)$, but relatively straight segments
with some undulations for the stiff melt ($\kappa=4)$. Interestingly,
we observe that entanglement points appear to cluster as the chain
stiffness is increased. This is consistent with the theory of polymer
knots\citep{metzler2002equilibrium,grosberg2007metastable}. Knots
on linear chains are tight because even though the bending energy
of a tight knot is higher than of a loose knot. This is due to the
maximization of conformational entropy which favors tight knots. The
topological constraints can also be rationalized as a tube around
the chain, where the tube diameter at the knot position is most narrow,
reflecting the local loss of conformational entropy. The tube diameter
in melts of long linear chains is constant along the chains, since
the entanglement density per length of chain is the same. This might
be different close to the chain ends. The clustering of entanglements
observed in Fig. \ref{fig:yearnmelt}, suggests that in analogy of
the knots, that the tubes confining the chains would be a co-existence
of narrow entanglement rich segments and wide entanglement sparse
segments. 

\subsection{Stress relaxation\label{sec:Stress-relaxation}}

\begin{figure}
\includegraphics[width=0.5\columnwidth]{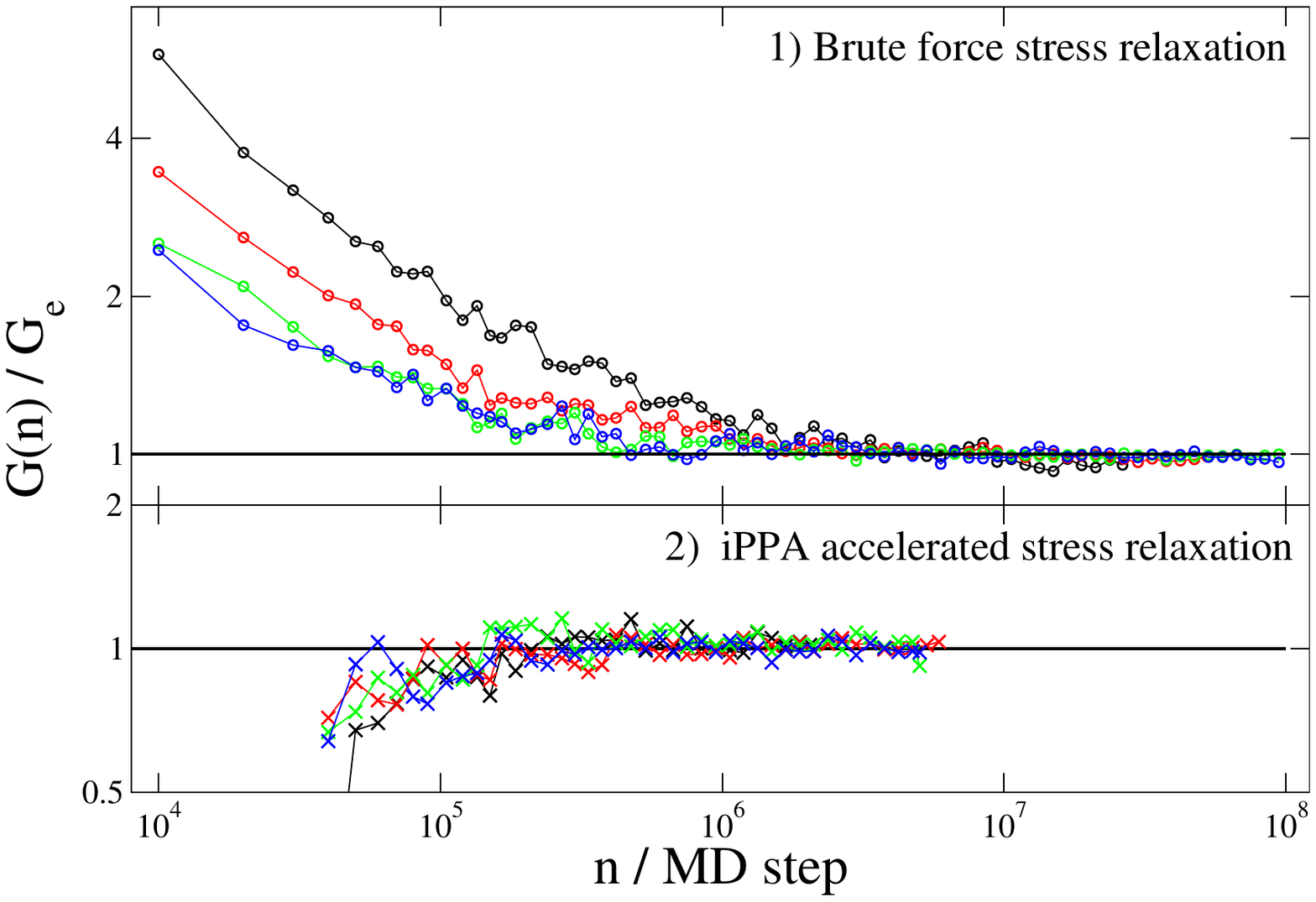}\caption{Reduced shear relaxation modulus of KG melts of stiffness values $\kappa=0,1,2,3$
(top plot with black, red, green, blue $\circ$, respectively) compared
to iPPA accelerated stress relaxation (bottom plot with similar colored
$\times$). \label{fig:Stress-relaxation}}
\end{figure}

We compare stress relaxation using two protocols 1) brute force KG
stress relaxation and 2) iPPA accelerated stress relaxation. To estimate
the equilibrium shear modulus, we use KG melts with approximately
$M=500$ chains and $Z=100$ entanglements per chain for stiffness
$\kappa=0,1,2$,3.\citep{SvaneborgEquilibration2022}{[}dataset{]}\citep{SvaneborgEquilibratedKGMeltsZ100}
The melts conformations were elongated by $\lambda=1.1$ ($10\%$
strain) during $10^{4}$ steps by stretching the simulation box along
the $x$ axis, while continually compressing it along the $yz$ directions
to keep the volume constant. We performed a long stress relaxation
simulation during which the average stress tensor was saved every
$10^{4}$ steps.

In the iPPA accelerated protocol, we start by performing PPA on the
melts to generate the corresponding meshes. Each mesh is deformed
using the same deformation protocol as the melt. To relax the mesh,
we ran a short simulation of $8\times10^{3}$ steps with the PPA force
field during which chain length can redistribute in the mesh and entanglements
can move to minimize the longitudinal tension. We then gradually use
the iPPA protocol to switch from the PPA to the KG force field during
a simulation of $10^{4}$ steps. Afterwards, we apply $2\times10^{3}$
MD steps to heat up the melt state to $T=1\epsilon$. Finally, we
proceed with a brute force KG stress relaxation as above. During the
force field transformations, we use a modified form of the FENE potential
in order to minimize topology violations, which will be discussed
below.

We estimate the time dependent shear relaxation modulus by

\[
G(n)=\frac{\sigma_{xx}(n)-\frac{1}{2}(\sigma_{yy}(n)+\sigma_{zz}(n))}{\lambda^{2}-\lambda^{-1}},
\]
where $\sigma_{\alpha\beta}(n)$ denotes the instantaneous microscopic
virial stress tensor. The equilibrium modulus is estimated as the
average of the brute force relaxed $G(n)$ for $n>2\times10^{6}$
steps. For the brute force simulation $n$ denotes the number of MD
steps of stress relaxation, while for a fair comparison we start counting
MD steps at the start of the mesh relaxation simulation. Thus the
first $2\times10^{4}$ steps of the iPPA accelerated protocol are
the computational cost of mesh relaxation, force field switching,
and the short warmup. During both protocols, chain ends are pinned
to avoid stress relaxation due to contour length fluctuations.\citep{DoiEdwards86} 

Fig. \ref{fig:Stress-relaxation} shows the time dependent shear relaxation
modulus for the two protocols. For the brute force stress relaxation,
we observe that stiffer melts reach their equilibrium modulus faster
than more flexible melts. We expect this fast relaxation takes of
the order of the entanglement time, and the entanglement time decreases
with increasing stiffness. We previously estimated the entanglement
time to be $\tau_{e}(\kappa=0)\approx1.3\times10^{6}$ and $\tau_{e}(\kappa=2)\approx9.3\times10^{4}$
steps, respectively.\citep{Svaneborg2020Characterization} We observe
that stress equilibrium is reached after roughly $2\times10^{6}$
steps for $\kappa=0$ and $0.6\times10^{6}$ for $\kappa=2$ in rough
agreement with the entanglement time. No reptation dynamics is possible
since the chain ends are pinned, hence the longest remaining relaxation
dynamics is equilibration of longitudinal tension along the chains,
which takes of the order of the Rouse time of a chain $\tau_{R}=Z^{2}\tau_{e}\sim10^{10}$
steps, which is clearly beyond the time scales that are feasible to
simulate.

For the iPPA accelerated protocol, we do not show data for the modulus
during the mesh relaxation and force field transformations, since
these are physically meaningless. The initial conformation in the
KG relaxation simulation is a deformed melt state, where the longitudinal
stresses have been equilibrated, but where the local chain structure
is perturbed due to the local wiggles created by the iPPA. We observe
that the shear modulus for all the different stiffnesses show the
same monotonous increasing behavior. After $10^{5}$ steps the stress
values of the iPPA accelerated protocol is within $20\%$ of the equilibrium
value for all systems, whereas the stress of the $\kappa=0$ system
is a factor two above the equilibrium for the brute force stress relaxation
protocol. After roughly $2\times10^{5}$ steps of the iPPA accelerated
protocol, all systems have reached the equilibrium shear moduli. Thus
we conclude that performing a PPA, deformation, fast relaxation, and
iPPA cycle before stress relaxation significantly reduce the time
it takes to reach the equilibrium stress response for $\kappa\lesssim2$,
whereas for stiffer melts $\kappa\gtrsim2$ the relaxation time of
the two protocols is comparable.

During the force field transformation, we continuously check for topology
violations. With the standard FENE potential, we observed a number
of topology violations for $\kappa\geq2$. We have tested a number
of variations of the deformation and relaxation protocol. The number
of topology violations depends strongly on stiffness, but also on
the number of steps taken during force field switching. As shown in
Fig. \ref{fig:Visualizations-of-topology-violations}, the chain crossing
transition state is one where two bonds are perpendicular to each
other. When a bond is stretched due to chain tension, the potential
barrier for topology violation is reduced. One can increase the potential
barrier by increasing the spring constant, however, this can cause
even more tension along the chains. Instead we replace FENE by a polynomial
expansion that for small bond extensions matches FENE, but grows faster
than FENE for larger bond extensions. This allows long bonds to be
penalized while not affecting the shorter bonds. We refer to the Appendix
for the details. This force field modification was observed to strongly
suppress topology violations, nonetheless we observe $2$ topology
violations for the $\kappa=2$, and $12$ topology violations for
the $\kappa=3$ simulations. Four of these are shown in Fig. \ref{fig:Visualizations-of-topology-violations}.
They all occur at $\lambda\approx0.3$ during the iPPA push-off, which
is where the chemical window $w(\lambda)$ jumps from $5$ to $4$
bonds distance. Since there are no computational advantages in using
the inverse primitive path protocol for the stiffer melts, we have
not investigated the topology violations further.

To summarize, we have shown that a PPA / deformation / relaxation
/ iPPA push-off can save us roughly an order of magnitude of stress
relaxation simulation time for melts with $\kappa\lesssim2$. For
the brute force stress relaxation dynamics local Brownian bead motion
equilibrates chain statistics on progressively larger and large scales.
This requires a very long simulation and typically necessitates access
to a super computer. The relaxation dynamics of the PPA mesh is energy
minimization, which very rapidly equilibrates the large scale mesh
structure. When excess contour length is reintroduced during the inverse
PPA back to the KG force field, the result is a strong, but as shown
above, local perturbation of the chain statistics due to the wiggles.
Thus the relevant relaxation time of the resulting melt is the Rouse
time of a single wiggle which is much shorter than the entanglement
time. Thus only a relative short simulation is required to establish
the equilibrium chain structure on all length scales.

\section{Conclusions\label{sec:Conclusions}}

Primitive path analysis\citep{PPA} (PPA) has been vastly successful
for the topological analysis of model polymer materials. The PPA algorithm
removes thermal fluctuations and generate the minimal tension mesh
of primitive paths, which allows the tube structure\citep{Edwards_tube_procphyssoc_67}
to be characterized, and enables the elastic properties of the material
due to entanglements to be inferred.\citep{sukumaran2005identifying,Gula2020Entanglement}
Here we have presented an iPPA algorithm, a force field transformation
that enables gradual reversible switching between the PPA and the
Kremer-Grest\citep{grest1986molecular} (KG) force fields. The effect
of iPPA is to slowly reintroduce the excess contour length in the
PPA meshes in a topology preserving way.

The effects of the iPPA force field transformation have been characterized,
and we have shown that $100$ cycles of PPA followed by iPPA preserves
the original melt topology. We have also illustrated how an iPPA push-off
can be used to convert synthetic PPA meshes into equilibrium KG model
melts. This enables detailed control over the topological state of
such model melts. As a toy example, we choose pseudo-2D knitted structures,
where entanglements form a regular 2D cubic lattice. These toy systems
hinted at complex physics and interactions between the entanglements,
especially for stiffer chains where we saw indications that entanglements
were forming clusters. This is consistent with expectations from theory
of confined polymer knots\citep{metzler2002equilibrium,grosberg2007metastable},
where maximization of conformational entropy causes knots to shrink
such that the topological constraint is localized.

Finally, we illustrated the utility combining PPA and iPPA to accelerate
stress relaxation. We deform a mesh rather than a melt, since relaxation
of the large scale mesh structure is essentially instantaneous since
it is just energy minimization. Then we use iPPA to convert the mesh
back to the KG force field, where we observed localized wiggles had
been created along the primitive path chains. Thus only a relatively
short simulation is required to reequilibrate local chain structure.
For melts with $\kappa\lesssim2$, we observed that the equilibrium
stress was reached roughly order of magnitude faster than brute force
stress relaxation. For stiffer systems, we observed topology violations
during iPPA, and no significant acceleration of the dynamics was observed
compared to brute force stress relaxation.

Here we have used melts as model systems, but we expect the iPPA push-off
to work equally well for KG model networks, where estimation of equilibrium
moduli also requires long simulations and careful extrapolation of
stress data.\citep{Gula2020Entanglement} iPPA also offers a way to
study the influence of topology of networks, we can e.g. relax frozen
entanglements with a phantom PPA analysis\citep{SGE_poly_05,Gula2020Entanglement},
and then convert a phantom mesh into a topologically relaxed KG melt.
Combining iPPA with thermodynamic integration techniques, we expect
the method enables the estimation of absolute free energies of deformed
model polymer materials. The iPPA force field has been implemented
in the Large Atomic Molecular Massively Parallel Simulator (LAMMPS)\citep{PlimptonLAMMPS,PlimptonLAMMPS2},
and is freely available from Ref. \citep{SvaneborgGITHUBippa} along
with all the scripts and input files required to reproduce the examples
presented here.

We acknowledge that part of the results of this research have been
achieved using the PRACE Research Infrastructure resource Joliot-Curie
SKL based in France at GENCI@CEA. Discussions with R. Everaers and
I.A. Gula are gratefully acknowledged.

\section*{Appendix\label{sec:Appendix}}

\subsection{Kremer-Grest model}

The KG polymer model\citep{grest1986molecular} is a bead-spring model
where all beads interact via WCA potential

\[
U_{WCA}(r)=\begin{cases}
4\epsilon\left[\left(\frac{\sigma}{r}\right)^{12}-\left(\frac{\sigma}{r}\right)^{6}+\frac{1}{4}\right] & r<r_{c}\\
0 & r\geq r_{c}
\end{cases}
\]
with $r_{c}=2^{1/6}\sigma$ being the minimum of the LJ potential.
In addition to the WCA interaction, bonded beads also interact via
the FENE potential

\[
U_{FENE}(r)=-\frac{kR^{2}}{2}\ln\left[1-\frac{r^{2}}{R^{2}}\right].
\]

Following Faller and Müller-Plathe\citep{faller1999local,faller2000local},
we augment KG the model by a angle dependent interaction potential 

\[
U(\Theta)=\kappa\epsilon\left(1-\cos\Theta\right),
\]
where $\Theta$ denotes the angle between subsequent bonds, and $\kappa$
is a dimensionless number controlling chain stiffness, this allows
the KG model to describe different species of chemical polymers,\citep{Svaneborg2020Characterization,Everaers2020Mapping}
Standard choices for the KG model are the parameters: $R=1.5\sigma$
and $k=30\epsilon/\sigma^{2}$. KG simulations are usually performed
at temperature $T=1\epsilon$ and at a bead density of $\rho_{b}=0.85\sigma^{-3}$.
This gives rise to an average bond length of $l_{b}=0.965\sigma$.
The standard thermostat chosen is a Langevin thermostat with friction
$\Gamma=0.5m_{b}\tau^{-1}$ where $\tau=\sigma\sqrt{m/\epsilon}$
defines the simulation unit of time, and $m$ denotes the bead mass.
We integrate the dynamics with a time step of $\Delta t=0.01\tau$
using the Farago/Grønbech-Jensen integrator using the Large Atomic
Molecular Massively Parallel Simulator (LAMMPS) code.\citep{gronbech2013simple,gronbech2014application,PlimptonLAMMPS,PlimptonLAMMPS2}

\subsection{Primitive-path analysis}

\begin{figure}
\includegraphics[width=0.5\columnwidth]{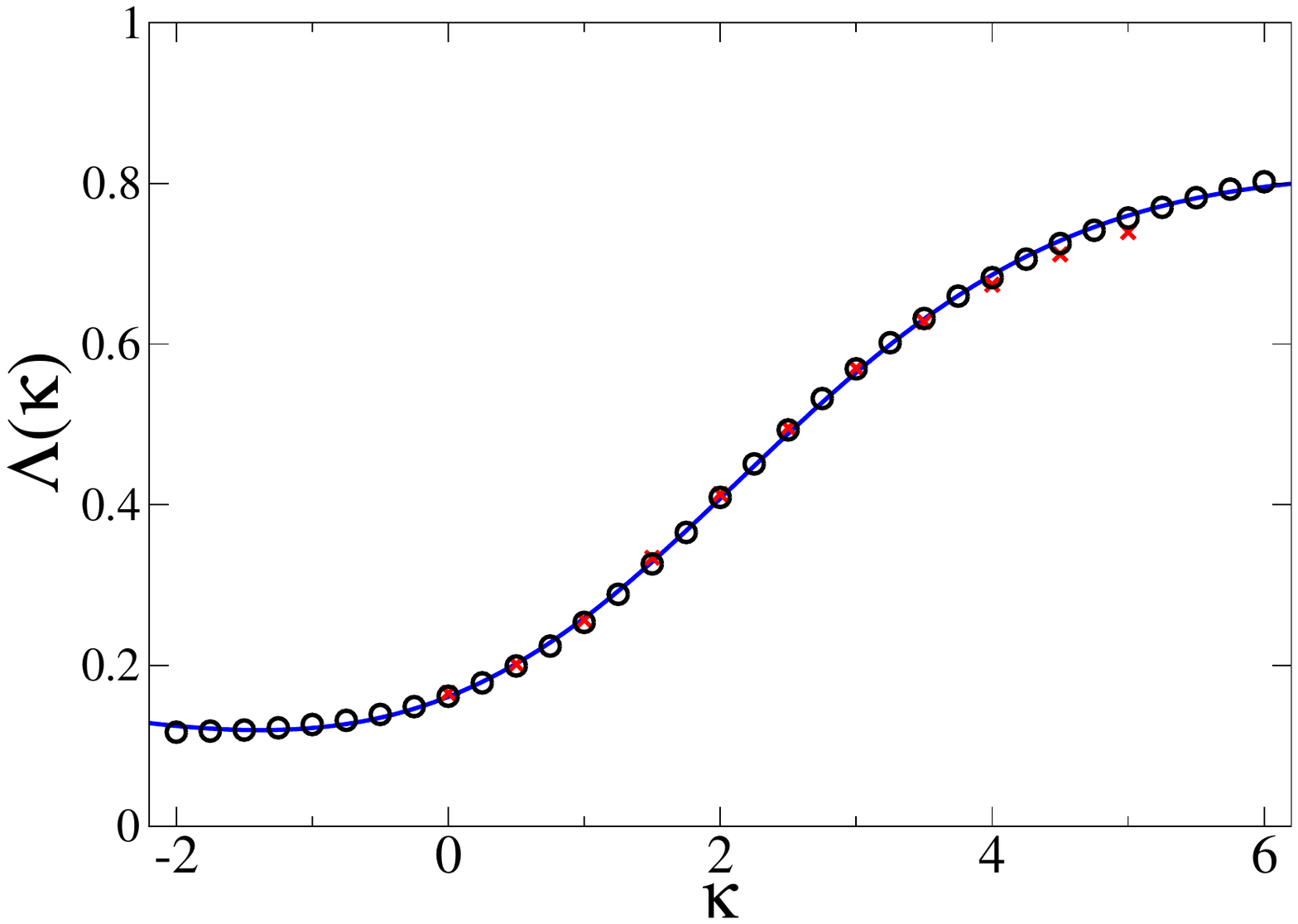}

\caption{Contour contraction as function of stiffness for primitive paths of
linear KG polymer melts with $Z=200$ entanglements per chain from
Ref. \citep{SvaneborgEquilibration2022} (black circles), and linear
KG melts with $N_{b}=400$ from Ref. \citep{dietz2022facile} (red
crosses) compared to an empirical Padé approximation $\Lambda(\kappa)=l_{pp}(\kappa)/l_{b}=(0.0273\kappa^{2}+0.0395\kappa+0.161)/(0.0486\kappa^{2}-0.169\kappa+1)$
(blue line). \label{fig:Contour-contraction}}
\end{figure}

The classical PPA algorithm\citep{PPA} proceeds as follows for a
melt: 1) the ends of all chains are fixed in space, 2) local intra
molecular pair interactions are disabled, and 3) the energy is minimized.
Since we retain inter molecular pair interactions different chains
are unable to pass through each other, thus preserving entanglements.
Energy minimization pulls the chains taut to minimize the bond energy
thus producing a unique primitive-path mesh. In the mesh chains are
approximately piecewise linear curves between entanglements, while
they are random walks in the melt state. During the PPA contraction
the excess length of the random walk conformations is lost, and the
minimal length is dictated by the density of entanglements. Thus we
can estimate the entanglement density from the degree of chain contraction,
and hence the plateau modulus.\citep{PPA,Svaneborg2020Characterization}. 

The classical PPA is performed by disabling all intramolecular WCA
interactions as well as the angular potential. The energy is minimizing
energy e.g. by damped Langevin dynamics. The spring constant is increased
to $k=100\epsilon\sigma^{-2}$ although this is an arbitrary number.
The temperature is chosen as $T=0.001\epsilon$, the friction is $\Gamma=20m\tau^{-1}$
and $10^{3}$ MD steps are performed with time step $\Delta t=0.006\tau$.
During this process chain length rapidly contracts as random-walk
like thermal fluctuations are converted to straight chain segments
between topological entanglements. To converge the process the friction
is reduced to $\Gamma=0.5m_{b}\tau^{-1}$ and relaxed for additional
$10^{5}$ steps. 

The result of the PPA analysis is the contour length contraction ratio
$\Lambda(\kappa)=l_{pp}(\kappa)/l_{b}$, where $l_{pp}$ denotes the
length of bonds after PPA. This ratio characterizes the density of
entanglements along the chain, and is usually converted into the number
of beads (or Kuhn segments) between entanglements. Fig, \ref{fig:Contour-contraction}
shows the contour length contraction for well-equilibrated highly
entangled linear KG polymer melts with varying stiffness from Ref.
\citep{SvaneborgEquilibration2022}{[}dataset{]}\citep{SvaneborgEquilibratedKGMeltsZ100}.
We observe a sigmoidal dependence of $\kappa$, where very flexible
chains show a large degree contraction, while stiff chains only show
a limited degree of contraction. This can be rationalized from the
diameter of the tube, flexible chains have wide tubes, stiff chains
narrow tubes. Hence there is significantly more excess length stored
in the thermal fluctuations of flexible chains compared to stiff chains.

\subsection{Modified FENE potential}

For the PPA force field, the transition state where topology preservation
is violated corresponds to a planar configuration with two perpendicular
bonds and four beads in a square arrangement.\citep{sukumaran2005identifying}
The equilibrium bond length is $1.22\sigma$. The energy of the transition
state can be increased by lowering $R$, however this reduces the
numerical stability of the simulation. To increase the energy of the
transition state we instead modified the Taylor expansion of the FENE
potential as

\[
U(r)=\frac{100\epsilon}{2}\left(\frac{r}{\sigma}\right)^{2}+100\epsilon\left(\frac{r}{\sigma}\right)^{6},
\]
for $r<0.5\sigma,$this is a very good approximation to the PPA FENE
potential, however for larger values it grows much faster than FENE.

\subsection{LAMMPS}

The PPA force field switching method described above has been implemented
as an extension to LAMMPS\citep{PlimptonLAMMPS,PlimptonLAMMPS2}.
For information on how to obtain the code, complete scripts, and examples
see Ref. \citep{SvaneborgGITHUBippa}. The following snippet of LAMMPS
code illustrates how to perform a simulation switching from the KG
to the PPA force field.
\begin{verbatim}
1: pair_style wca/ppa window 10 u0 200 alpha 2.32 lambda 1.0
2: pair_coeff 1*2 1*2 1.0 1.0
3: bond_style fene
4: bond_coeff  *   100 1.5 0.0 1.0
5: special_bonds  lj 1 1 1
6: variable switch_to_ppa equal "ramp(1.00,0.00)"
7: fix switch all adapt  1  pair wca/ppa lambda 1*2 1*2 v_switch_to_ppa 
8: run 10000

\end{verbatim}
The first line defines the protocol of the force field switch. The
key words specify parameters that match the notation of the present
paper. The last argument ``lambda 1.0'' sets the initial value of
$\lambda$, for instance if we wanted to run a simulation with a constant
preset value of lambda. The second line sets up the specific interactions
assuming a system with two types of beads with WCA parameters $\epsilon=1.0$
and $\sigma=1.0$. The third line defines the bond potential as FENE.
The fourth line specifies the FENE bond parameters spring constant
of $k=100\epsilon\sigma^{-2}$, cutoff distance $R=1.5\sigma$, finally
a non-standard value of $\epsilon=0.0$ and $\sigma=1.0$ between
bonded beads. The fifth line specifies that WCA/PPA pair interactions
should be calculated between nearest, next-nearest, and second-next-nearest
neighbors. The usual LAMMPS convention is to include the WCA interaction
between bonded beads as part of a FENE+WCA bond potential. Then pair
interactions should not be calculated again for bonded beads. However,
here we need to explicitly use the wca/ppa forcefield for all pair
interactions, hence we disabled the WCA contribution to the FENE+WCA
bond potential in line four, and specify that we want a full pair
interaction to be calculated between bonded beads in line 5. The sixth
line defines a variable which changes as a linear ramp from the value
of $1.0$ at the start of the run to $0.0$ at the end of the run.
The sevenths line sets up a fix that at every time step changes the
``lambda'' parameter of the wca/ppa interaction with the instantaneous
value. Finally, the eighth line runs a simulation with 10000 integration
steps performing the force field transformation. The reverse transformation
is simply obtained by changing the ramp and the initial value of lambda
in the force field. In case of circular polymer chains, the wca/ppa
takes an optional argument ``circular'', which changes the definition
of chemical distance to that of circular chains. Note that additional
commands are required to load the mesh, setup integration and thermostats
and pin the chain ends.

To limit topology violations, the FENE potential can be replaced by
a polynomial expansion. That is achieved by replacing line 3-4 above
by the following
\begin{verbatim}
3: bond_style poly06
4: bond_coeff      *  0.0     0.0  0.0  50.0 0.0 0.0  0.0   100.0
\end{verbatim}
Where the arguments after the star is the point around which we perform
the Taylor expansion, and then the $7$ coefficients defining a 6th
degree polynomial expansion of the potential. 

The following snippet of LAMMPS code sets up a switching force field
with concurrent topology checking. Topology is checked via a dummy
pair style associated with the dummy beads that represent the neighbor
list of bonds. This does not add forces, but only checks bond pairs
for topology violations and stores these in a separate fix ``f\_topo''
that is defined automatically by the pair style.
\begin{verbatim}
1: pair_style      hybrid  wca/ppa window 10 alpha 2.32 lambda  1.0  u0 200 \
2:                         topo    window 10 alpha 2.32 lambda  1.0  cutoff 2.0  bondtype *  
3: pair_coeff      1*2  1*2  wca/ppa  1.0 1.0
4: pair_coeff      1*2   3   none
5: pair_coeff      3     3   topo
6: fix iPPA all adapt  1  pair wca/ppa lambda 1*2 1*2 v_switch_to_ppa \
7:                        pair topo    lambda  3   3  v_switch_to_ppa 

\end{verbatim}
Here we assume the input configuration has two bead types 1 and 2
(e.g. to distinguish free beads from. fixed chain ends). The topology
check adds a third bead type which are the type 3 beads placed at
the center of bonds. Line 1 and 2 is one command that sets up a hybrid
force field using the ``wca/ppa'' force field defined above as well
as the dummy force field for checking for topology violations ``topo''.
``cutoff 2.0'' sets up the cutoff distance for checking for topology
violations. The check is applied to all bond types ``bondtype {*}''.
Since intramolecular bond pairs within the current switching window
should not be included in the check, we need again to specify how
we switch the chemical window. The third line defines the WCA interactions
$\epsilon=1.0$ and $\sigma=1.0$ between all beads of type 1 and
2. The fourth line defines the cross-interaction between beads and
dummy beads as none. Since the latter are a computational trick to
keep track of bonds. The fifth line sets the ``pair interaction''
between pairs of dummy beads (representing bonds) as the topo pair
style. The final sixth line sends the current value of the switching
variable to the two pair styles, since they both need it to calculate
the current window of chemical distances $w(\lambda)$.

When this pair style is used, a fix denoted ``f\_topo'' in LAMMPS
is automatically created when the pair style is initialized, and it
exports four global scalars f\_topo{[}1{]} to f\_topo{[}4{]} which
can be output by LAMMPS. These four scalars are 1) the total number
of topology violations identified during the current time step, 2)
the accumulated number of topology violations so far, 3) the total
number of intermolecular topology violations during the current time
step, and 4) the accumulated number of inter-molecular topology violations.
To flip beads back to reverse topology violations an optional argument
``flip'' can be given to the topo pair style. In the case of circular
chains, the optional argument ``circular'' must be given both to
the ``wca/ppa'' and ``topo'' pair interactions to make sure they
apply the same intramolecular forces and topology violation checks.


\begin{thebibliography}{10}
\expandafter\ifx\csname url\endcsname\relax
  \def\url#1{\texttt{#1}}\fi
\expandafter\ifx\csname urlprefix\endcsname\relax\def\urlprefix{URL }\fi
\expandafter\ifx\csname href\endcsname\relax
  \def\href#1#2{#2} \def\path#1{#1}\fi

\bibitem{DoiEdwards86}
M.~Doi, S.~F. Edwards, The Theory of Polymer Dynamics, Clarendon, Oxford, 1986.

\bibitem{Edwards_procphyssoc_67}
S.~F. Edwards, Statistical mechanics with topological constraints: I, Proc.
  Phys. Soc. 91 (1967) 513.

\bibitem{degennes71}
P.~G. de~Gennes, Reptation of a polymer chain in the presence of fixed
  obstacles, J. Chem. Phys. 55 (1971) 572.

\bibitem{muller1996topological}
M.~M{\"u}ller, J.~Wittmer, M.~Cates, Topological effects in ring polymers: A
  computer simulation study, Phys. Rev. E 53 (1996) 5063.

\bibitem{grosberg2014annealed}
A.~Y. Grosberg, Annealed lattice animal model and {F}lory theory for the melt
  of non-concatenated rings: {T}owards the physics of crumpling, Soft Matter 10
  (2014) 560.

\bibitem{rosa2014ring}
A.~Rosa, R.~Everaers, Ring polymers in the melt state: The physics of
  crumpling, Phys. Rev. Lett. 112 (2014) 118302.

\bibitem{kapnistos2008unexpected}
M.~Kapnistos, M.~Lang, D.~Vlassopoulos, W.~Pyckhout-Hintzen, D.~Richter,
  D.~Cho, T.~Chang, M.~Rubinstein, Unexpected power-law stress relaxation of
  entangled ring polymers, Nat. Mater. 7 (2008) 997.

\bibitem{ge2016self}
T.~Ge, S.~Panyukov, M.~Rubinstein, Self-similar conformations and dynamics in
  entangled melts and solutions of nonconcatenated ring polymers,
  Macromolecules 49 (2016) 708.

\bibitem{huang2019unexpected}
Q.~Huang, J.~Ahn, D.~Parisi, T.~Chang, O.~Hassager, S.~Panyukov, M.~Rubinstein,
  D.~Vlassopoulos, Unexpected stretching of entangled ring macromolecules,
  Phys. Rev. Lett. 122 (2019) 208001.

\bibitem{grosberg1993crumpled}
A.~Grosberg, Y.~Rabin, S.~Havlin, A.~Neer, Crumpled globule model of the
  three-dimensional structure of {DNA}, Europhys. Lett. 23 (1993) 373.

\bibitem{rosa2008structure}
A.~Rosa, R.~Everaers, Structure and dynamics of interphase chromosomes, PLoS.
  Comput. Biol. 4 (2008) e1000153.

\bibitem{rosa2010looping}
A.~Rosa, N.~B. Becker, R.~Everaers, Looping probabilities in model interphase
  chromosomes, Biophys. J. 98 (2010) 2410.

\bibitem{chen1995topology}
J.~Chen, C.~A. Rauch, J.~H. White, P.~T. Englund, N.~R. Cozzarelli, The
  topology of the kinetoplast {DNA} network, Cell 80 (1995) 61.

\bibitem{klotz2020equilibrium}
A.~R. Klotz, B.~W. Soh, P.~S. Doyle, Equilibrium structure and deformation
  response of 2{D} kinetoplast sheets, PNAS 117 (2020) 121.

\bibitem{dietrich1983nouvelle}
C.~O. Dietrich-Buchecker, J.~Sauvage, J.~Kintzinger, Une nouvelle famille de
  molecules: les metallo-catenanes, Tetrahedron Lett. 24 (1983) 5095.

\bibitem{thompson1964reactions}
M.~C. Thompson, D.~H. Busch, Reactions of coordinated ligands. {IX}.
  utilization of the template hypothesis to synthesize macrocyclic ligands in
  situ, JACS 86 (1964) 3651.

\bibitem{busch1992structural}
D.~H. Busch, Structural definition of chemical templates and the prediction of
  new and unusual materials, J. Inclus. Phenom. Mol. Recognit. Chem. 12 (1992)
  389.

\bibitem{hubin2000template}
T.~J. Hubin, D.~H. Busch, Template routes to interlocked molecular structures
  and orderly molecular entanglements, Coord. Chem. Rev. 200 (2000) 5.

\bibitem{wu2017poly}
Q.~Wu, P.~M. Rauscher, X.~Lang, R.~J. Wojtecki, J.~J. De~Pablo, M.~J. Hore,
  S.~J. Rowan, Poly [n] catenanes: Synthesis of molecular interlocked chains,
  Science 358 (2017) 1434.

\bibitem{datta2020self}
S.~e.~a. Datta, Self-assembled poly-catenanes from supramolecular toroidal
  building blocks, Nature 583 (2020).

\bibitem{dietrich1989synthetic}
C.~O. Dietrich-Buchecker, J.-P. Sauvage, A synthetic molecular trefoil knot,
  Angew. Chem. 28 (1989) 189.

\bibitem{segawa2019topological}
Y.~Segawa, M.~Kuwayama, Y.~Hijikata, M.~Fushimi, T.~Nishihara, J.~Pirillo,
  J.~Shirasaki, N.~Kubota, K.~Itami, Topological molecular nanocarbons:
  All-benzene catenane and trefoil knot, Science 365 (2019) 272.

\bibitem{li2022robust}
G.~Li, J.~Zhao, Z.~Zhang, X.~Zhao, L.~Cheng, Y.~Liu, Z.~Guo, W.~Yu, X.~Yan,
  Robust and dynamic polymer networks enabled by woven crosslinks, Angew. Chem.
  134 (2022) e202210078.

\bibitem{orlandini2021topological}
E.~Orlandini, C.~Micheletti, Topological and physical links in soft matter
  systems, J. Condens. Matter Phys. 34 (2021) 013002.

\bibitem{zhang2022molecular}
Z.-H. Zhang, B.~J. Andreassen, D.~P. August, D.~A. Leigh, L.~Zhang, Molecular
  weaving, Nat. Mater. 21 (2022) 275.

\bibitem{ashbridge2022knotting}
Z.~Ashbridge, S.~D. Fielden, D.~A. Leigh, L.~Pirvu, F.~Schaufelberger,
  L.~Zhang, Knotting matters: orderly molecular entanglements, Chem. Soc. Rev.
  (2022).

\bibitem{binder1995monte}
K.~Binder, Monte Carlo and Molecular Dynamics simulations in polymer science,
  Oxford University Press: New York, 1995.

\bibitem{kremer2000computer}
K.~Kremer, Computer simulations in soft matter science, in: M.~E. Cates, M.~R.
  Evans (Eds.), Soft and fragile matter: Non equilibrium dynamics,
  metastability and flow, J. W. Arrowsmith Ltd., Bristol, U.K., 2000, p. 145.

\bibitem{MultiscalePeterKremerFaradayDiss2010}
C.~Peter, K.~Kremer, Multiscale simulation of soft matter systems, Faraday
  Discuss. 144 (2010) 9.

\bibitem{grest1986molecular}
G.~S. Grest, K.~Kremer, Molecular dynamics simulation for polymers in the
  presence of a heat bath, Phys. Rev. A 33 (1986) 3628.

\bibitem{kremer90}
K.~Kremer, G.~S. Grest, Dynamics of entangled linear polymer melts: A molecular
  dynamics simulation, J. Chem. Phys. 92 (1990) 5057.

\bibitem{grest90a}
G.~S. Grest, K.~Kremer, Statistical properties of random cross-linked rubbers,
  Macromolecules 23 (1990) 4994.

\bibitem{halverson2011molecular}
J.~D. Halverson, W.~B. Lee, G.~S. Grest, A.~Y. Grosberg, K.~Kremer, Molecular
  dynamics simulation study of nonconcatenated ring polymers in a melt. {I}.
  {S}tatics, J. Chem. Phys. 134 (2011) 204904.

\bibitem{schram2019local}
R.~D. Schram, A.~Rosa, R.~Everaers, Local loop opening in untangled ring
  polymer melts: {A} detailed "{F}eynman test" of models for the large scale
  structure, Soft Matter 15 (2019) 2418.

\bibitem{halverson2012}
J.~D. Halverson, G.~S. Grest, A.~Y. Grosberg, K.~Kremer, Rheology of ring
  polymer melts: {F}rom linear contaminants to ring-linear blends, Phys. Rev.
  Lett. 108 (2012) 038301.

\bibitem{michieletto2020dynamical}
D.~Michieletto, T.~Sakaue, Dynamical entanglement and cooperative dynamics in
  entangled solutions of ring and linear polymers, ACS Macro Lett. 10 (2020)
  129.

\bibitem{tubiana2022circular}
L.~Tubiana, F.~Ferrari, E.~Orlandini, Circular polycatenanes: {S}upramolecular
  structures with topologically tunable properties, Phys. Rev. Lett. 129 (2022)
  227801.

\bibitem{caraglio2015stretching}
M.~Caraglio, C.~Micheletti, E.~Orlandini, Stretching response of knotted and
  unknotted polymer chains, Phys. Rev. Lett. 115 (2015) 188301.

\bibitem{Gula2020Entanglement}
I.~A. Gula, H.~A. Karimi-Varzaneh, C.~Svaneborg, Computational study of
  cross-link and entanglement contributions to the elastic properties of model
  {PDMS} networks, Macromolecules 53 (2020) 6907.

\bibitem{PPA}
R.~Everaers, S.~K. Sukumaran, G.~S. Grest, C.~Svaneborg, A.~Sivasubramanian,
  K.~Kremer, Rheology and microscopic topology of entangled polymeric liquids,
  Science 303 (2004) 823.

\bibitem{sukumaran2005identifying}
S.~K. Sukumaran, G.~S. Grest, K.~Kremer, R.~Everaers, Identifying the primitive
  path mesh in entangled polymer liquids, J. Polym. Sci., Part B: Polym. Phys.
  43 (2005) 917.

\bibitem{Svaneborg2020Characterization}
C.~Svaneborg, R.~Everaers, Characteristic time and length scales in melts of
  {K}remer-{G}rest bead-spring polymers with wormlike bending stiffness,
  Macromolecules 53 (2020) 1917.

\bibitem{kroger2005shortest}
M.~Kr{\"o}ger, Shortest multiple disconnected path for the analysis of
  entanglements in two-and three-dimensional polymeric systems, Comput. Phys.
  Commun. 168 (2005) 209.

\bibitem{kroger2023z1plus}
M.~Kr{\"o}ger, J.~D. Dietz, R.~S. Hoy, C.~Luap, The {Z}1+ package: Shortest
  multiple disconnected path for the analysis of entanglements in
  macromolecular systems, Comput. Phys. Commun. 283 (2023) 108567.

\bibitem{tzoumanekas2006topological}
C.~Tzoumanekas, D.~N. Theodorou, Topological analysis of linear polymer melts:
  {A} statistical approach, Macromolecules 39 (2006) 4592.

\bibitem{shanbhag2005chain}
S.~Shanbhag, R.~G. Larson, Chain retraction potential in a fixed entanglement
  network, Phys. Rev. Lett. 94 (2005) 076001.

\bibitem{zhou2005primitive}
Q.~Zhou, R.~G. Larson, Primitive path identification and statistics in
  molecular dynamics simulations of entangled polymer melts, Macromolecules 38
  (2005) 5761.

\bibitem{everaers2012topological}
R.~Everaers, Topological versus rheological entanglement length in
  primitive-path analysis protocols, tube models, and slip-link models, Phys.
  Rev. E. 86 (2012) 022801.

\bibitem{zhang2014equilibration}
G.~Zhang, L.~A. Moreira, T.~Stuehn, K.~C. Daoulas, K.~Kremer, Equilibration of
  high molecular weight polymer melts: {A} hierarchical strategy, ACS Macro.
  Lett. 3 (2014) 198.

\bibitem{moreira2015direct}
L.~A. Moreira, G.~Zhang, F.~M{\"u}ller, T.~Stuehn, K.~Kremer, Direct
  equilibration and characterization of polymer melts for computer simulations,
  Macromol. Theory Simul. 24 (2015) 419.

\bibitem{SvaneborgEquilibration2016}
C.~Svaneborg, H.~A. Karimi-Varzaneh, N.~Hojdis, F.~Fleck, R.~Everaers,
  Multiscale approach to equilibrating model polymer melts, Phys. Rev. E 94
  (2016) 032502.

\bibitem{SvaneborgEquilibration2022}
C.~Svaneborg, R.~Everaers, Multiscale equilibration of highly entangled
  isotropic model polymer melts., J. Chem. Phys. 158 (2023) 054903.

\bibitem{PlimptonLAMMPS}
S.~Plimpton, Fast parallel algorithms for short-range molecular dynamics, J.
  Comp. Phys. 117 (1995) 1.

\bibitem{PlimptonLAMMPS2}
A.~P.~T. et~al., {LAMMPS} a flexible simulation tool for particle based
  materials modeling at the atomic, meso, and continuum scales, Comput. Phys.
  Commun. 271 (2022) 108171.

\bibitem{sirk2012enhanced}
T.~W. Sirk, Y.~R. Slizoberg, J.~K. Brennan, M.~Lisal, J.~W. Andzelm, An
  enhanced entangled polymer model for dissipative particle dynamics, J. Chem.
  Phys. 136 (2012) 134903.

\bibitem{kcrane}
K.~Crane, \href{https://github.com/keenancrane/plain-knit-yarn}{{A simple
  parametric model of plain-knit yarns}} (March 2023).
\newline\urlprefix\url{https://github.com/keenancrane/plain-knit-yarn}

\bibitem{metzler2002equilibrium}
R.~Metzler, A.~Hanke, P.~G. Dommersnes, Y.~Kantor, M.~Kardar, Equilibrium
  shapes of flat knosvats, Phys. Rev. Lett. 88 (2002) 188101.

\bibitem{grosberg2007metastable}
A.~Y. Grosberg, Y.~Rabin, Metastable tight knots in a wormlike polymer, Phys.
  Rev. Lett. 99 (2007) 217801.

\bibitem{SvaneborgEquilibratedKGMeltsZ100}
C.~Svaneborg, R.~Everaers,
  \href{https://doi.org/10.5281/zenodo.7319837}{Equilibrated {K}remer-{G}rest
  polymer melts of {M}=500 linear chains with {Z}=100 entanglements for varying
  chain stiffness.} (Feb. 2023).
\newblock \href {https://doi.org/10.5281/zenodo.7319837}
  {\path{doi:10.5281/zenodo.7319837}}.
\newline\urlprefix\url{https://doi.org/10.5281/zenodo.7319837}

\bibitem{Edwards_tube_procphyssoc_67}
S.~F. Edwards, The statistical mechanics of polymerized material, Proc. Phys.
  Soc. 92 (1967) 9.

\bibitem{SGE_poly_05}
C.~Svaneborg, G.~S. Grest, R.~Everaers, Disorder effects on the strain response
  of model polymer networks, Polymer 46 (2005) 4283.

\bibitem{SvaneborgGITHUBippa}
C.~Svaneborg, Lammps implementation of inverse {PPA} analysis
  https://github.com/zqex/i{PPA} (2024).

\bibitem{faller1999local}
R.~Faller, A.~Kolb, F.~M{\"u}ller-Plathe, Local chain ordering in amorphous
  polymer melts: Influence of chain stiffness, Phys. Chem. Chem. Phys. 1 (1999)
  2071.

\bibitem{faller2000local}
R.~Faller, F.~M{\"u}ller-Plathe, A.~Heuer, Local reorientation dynamics of
  semiflexible polymers in the melt, Macromolecules 33 (2000) 6602.

\bibitem{Everaers2020Mapping}
R.~Everaers, H.~A. Karimi-Varzaneh, F.~Fleck, N.~Hojdis, C.~Svaneborg,
  {K}remer-{G}rest models for commodity polymer melts: {L}inking theory,
  experiment, and simulation at the {K}uhn scale, Macromolecules 53 (2020)
  1901.

\bibitem{gronbech2013simple}
N.~Gr{\o}nbech-Jensen, O.~Farago, A simple and effective {V}erlet-type
  algorithm for simulating {L}angevin dynamics, Mol. Phys. 111 (2013) 983.

\bibitem{gronbech2014application}
N.~Gr{\o}nbech-Jensen, N.~R. Hayre, O.~Farago, Application of the {G-JF}
  discrete-time thermostat for fast and accurate molecular simulations, Comput.
  Phys. Commun. 185 (2014) 524.

\bibitem{dietz2022facile}
J.~D. Dietz, R.~S. Hoy, Facile equilibration of well-entangled semiflexible
  bead--spring polymer melts, J. Chem. Phys. 156 (2022) 014103.

\end{thebibliography}
\end{document}